\documentclass[a4paper,11pt,twoside]{article}
\pdfoutput=1

\usepackage{geometry} 
\usepackage{amssymb,amsmath,bm}
\usepackage{color}
\usepackage{slashed}
\usepackage{graphicx}
\usepackage[latin1]{inputenc}
\usepackage{amsfonts}
\usepackage{lscape}
\usepackage{amsthm}
\usepackage{booktabs}
\usepackage{array}
\usepackage{rotating}
\usepackage{float}

\usepackage[small,bf]{caption}
\selectfont

\newcommand{\nt}{\widetilde{\chi}^0}
\newcommand{\ch}{\widetilde{\chi}^\pm}
\newcommand{\chp}{\widetilde{\chi}^+}
\newcommand{\chm}{\widetilde{\chi}^-}
\newcommand{\stau}{\widetilde{\tau}}
\newcommand{\ord}[1]{\mathcal{O}\left( #1 \right)}

\newcommand{\prospino}[0]{\text{\tt Prospino 2}}

\newcommand{\herwig}[0]{\text{\tt Herwig++}}
\newcommand{\hepmc}[0]{\text{\tt HepMC}}
\newcommand{\delphes}[0]{\text{\tt Delphes 3}}

\def\gev{\ensuremath{~\text{GeV}}}

\newcommand{\fbai}{\unskip\,{\text{fb}^{-1}}}
\def\missingET{\ensuremath{\displaystyle{\not}E_T}}
\newcommand{\pT}{\ensuremath{p^\mathrm{T}}}
\def\GeV{\ensuremath{\,\text{GeV}}}
\def\TeV{\ensuremath{\,\text{TeV}}}

\begin{document}
\begin{titlepage}

\vspace*{-15mm}
\begin{flushright}
ULB-TH/13-10, MPP-2013-162, STUPP-2013-213
\end{flushright}
\vspace*{0.7cm}

\begin{center}

{
\bf\LARGE
Cornering light Neutralino Dark Matter at the LHC}
\\[8mm]
Lorenzo~Calibbi $^{\star}$\footnote{E-mail: \texttt{lcalibbi@ulb.ac.be}},
Jonas~M.~Lindert $^{\dagger}$\footnote{E-mail: \texttt{lindert@mppmu.mpg.de}},
Toshihiko~Ota $^{\ddag}$\footnote{E-mail: \texttt{toshi@mppmu.mpg.de}},
Yasutaka~Takanishi $^{*}$\footnote{E-mail: \texttt{yasutaka@mpi-hd.mpg.de}}
\\[1mm]
\end{center}
\vspace*{0.50cm}
\centerline{$^{\star}$ \it
 Service de Physique Th\'eorique, Universit\'e Libre de Bruxelles,}
\centerline{\it
Bld du Triomphe, CP225, B-1050 Brussels, Belgium}
\vspace*{0.2cm}
\centerline{$^{\dagger}$ \it
Max-Planck-Institut f\"ur Physik (Werner-Heisenberg-Institut),}
\centerline{\it
F\"ohringer Ring 6, D-80805 M\"unchen, Germany}
\vspace*{0.2cm}
\centerline{$^{\ddag}$ \it
Department of Physics, Saitama University,}
\centerline{\it
Shimo-Okubo 255, 338-8570 Saitama-Sakura, Japan}
\vspace*{0.2cm}
\centerline{$^*$ \it
Max-Planck-Institut f\"ur Kernphysik,}
\centerline{\it
Saupfercheckweg 1, D-69117 Heidelberg, Germany}

\vspace*{1.20cm}
\begin{abstract}
\noindent
We investigate the current status of the light neutralino dark
matter scenario within the minimal supersymmetric standard model
(MSSM) taking into account latest results from the LHC.
A discussion of the relevant constraints, in particular from the
dark matter relic abundance, leads us to a manageable simplified 
model defined by a subset of MSSM parameters. Within this simplified
model we reinterpret a recent search for electroweak supersymmetric particle
production based on a signature including multi-taus plus missing transverse 
momentum performed by the ATLAS collaboration. In this way we
derive stringent constraints on the light neutralino parameter space. 
In combination with further experimental information from the LHC,
such as dark matter searches in the monojet channel and constraints on
invisible Higgs decays, we obtain a lower bound on the lightest
neutralino mass of about \mbox{24 GeV}.
This limit is stronger than any current limit set by underground direct
dark matter searches or indirect detection experiments. 
With a mild improvement of the sensitivity of the multi-tau search,
light neutralino dark matter can be fully tested up to about 
\mbox{30 GeV}.
\end{abstract}

\end{titlepage}

%%%%%%%%%%%%%%%%%%%%%%%%%%%%%%%%%%%%%%%%%
\section{Introduction}
\addtocounter{footnote}{-4}

Searches for supersymmetry (SUSY) performed by the Large Hadron
Collider (LHC) experiments on the center-of-mass energy $\sqrt{s}=7,\,8$ TeV run data have
started to set stringent constraints to the strongly-interacting SUSY
particles. They translate to limits on the SUSY masses up to $1\div
1.5$ TeV for gluinos and squarks of the first family
\cite{ATLAS:2012ona} and $600\div 700$ GeV for third generation
squarks \cite{ATLAS:2013cma}. Comparatively weaker constraints were
obtained for the states that can be only produced through electroweak
(EW) interactions, such as sleptons, EW gauginos and Higgsinos. The
resulting bounds are up to $200\div 300$ GeV for selectrons and smuons
\cite{CMS:aro} and up to $600$ GeV for charginos
\cite{ATLAS:2013rla}. They often depend on the assumption of available
on-shell decays.  Nevertheless, it is remarkable that the LHC
experiments have started -- especially with the 2012 run data at
$\sqrt{s}=8$ TeV -- to go considerably beyond LEP in testing the EW
sector of the theory.  Such direct searches for EW production of SUSY
particles are of major importance, as the EW-interacting particles can
be in principle much lighter than the strongly-interacting ones.

In this paper, we want to discuss how the 8 TeV run data constrain the
parameter space that is compatible with a light neutralino as
candidate for cold dark matter (DM). In other words, we aim at
answering the following question: how light can the lightest
neutralino still be after the 8 TeV run of the LHC?\footnote{For early
  works addressing this question, see
  e.g.~\cite{Hooper:2002nq,Belanger:2002nr,Bottino:2002ry}.}

The framework we adopt for our study can be defined as follows:
\begin{itemize}
\item only the field content of the MSSM is considered;
\item DM is a thermal relic and a standard history of the universe is
  assumed;
\item the abundance of the lightest MSSM neutralino, whose stability is guaranteed by R-parity
  conservation, is required to not exceed the observed DM relic abundance.
\end{itemize}
We are not going to make any assumption on the origin and the
relations among the SUSY-breaking parameters: in particular we drop
the hypothesis of gaugino mass unification that would imply the lower
bound for the lightest neutralino mass reported by the PDG \cite{PDG}:
$m_{\widetilde{\chi}^{0}_{1}} > 46~\text{GeV}$. Instead, we treat all
SUSY soft-masses as free low-energy parameters.

As we discuss in the next section, relic density constraints from
cosmic microwave background (CMB)
observations identify the parameter space compatible with light
neutralino dark matter and thus the features of the SUSY spectrum and
the LHC phenomenology.  In particular, we are going to argue that,
under the assumptions listed above, neutralino DM with
\begin{equation}
m_{\widetilde{\chi}^{0}_{1}} \lesssim 30~\text{GeV}
\label{eq:mchi30}
\end{equation}
is only possible in a specific region of the supersymmetric parameter
space, featuring relatively light staus and Higgsinos. Hence, searches
based on multi-tau plus missing energy events are particularly
promising in order to test such a scenario, as we are going to discuss
in detail. A similar study has been presented last year in
Ref.~\cite{Belanger:2012jn}. Here, we want to generalize the approach
of Ref.~\cite{Belanger:2012jn}, by considering a simplified model
defined only by the subset of SUSY parameters relevant for the
determination of the relic abundance. The rest of the spectrum is
allowed to be heavy, so that our study is very conservative as it
considers only EW production of staus and neutral and charged
Higgsinos.  Moreover, the ATLAS collaboration has recently performed a
search for new physics in a final state with at least two taus and
large missing transverse momentum \cite{ATLAS:2013yla}, that we will
translate into a stringent test of the light neutralino parameter
space.\footnote{After the completion of this work, a similar analysis, 
based on $19.5~\fbai$ at $\sqrt{s}=8$ TeV, has been presented by CMS~\cite{new-CMS}. The
exclusion limit they obtain seems to be perfectly compatible with the results of Ref.~\cite{ATLAS:2013yla}.}

The rest of this paper is organized as follows. In section
\ref{sec:lightDM} we qualitatively review the features of the
parameter space compatible with light neutralino DM and we define the
simplified model we are going to employ in the rest of the study. A
quantitative discussion of the relevant constraints and the result of
a numerical scan of the parameter space are presented in section
\ref{sec:relicdensity}, while bounds from direct and indirect dark
matter searches are briefly discussed in \ref{sec:DMsearches}. Section
\ref{sec:LHC_pheno} is dedicated to a discussion of the spectrum of
the light DM scenario and the consequences for SUSY searches at the
LHC. In section \ref{sec:otherLHC}, we discuss the indirect limits on
the parameter space from invisible Higgs decays and we show the possible impact
of direct DM searches at the LHC in the monojet plus missing energy
channel. In section \ref{sec:LHC_numerics}, we present a Monte Carlo
study reproducing the limits from
Ref.~\cite{ATLAS:2013yla}. Afterwards we translate these limits into
bounds on the light neutralino parameter space.  We conclude
summarizing our results in section \ref{sec:conclusions}.

%%%%%%%%%%%%%%%%%%%%%%%%%%%%%%%%%%%%%%%%%%%%%%%%%%%%%%%
\section{Light neutralino DM in the MSSM}
\label{sec:lightDM}
In this section, we discuss how the parameter space of the MSSM is
constrained by the requirement of neutralino dark matter with
$m_{\widetilde{\chi}^{0}_{1}} \lesssim 30$ GeV.
The first obvious condition is that, due to the LEP bound on the
chargino mass \cite{PDG} that implies for the wino and Higgsino mass
parameters $M_2,\,\mu\gtrsim 90$ GeV, the lightest neutralino is
mainly bino-like with the bino mass $M_1$ approximately:
\begin{align}
M_{1} \lesssim 30~\text{GeV}.
\label{eq:Our-parameter-region-Nr1}
\end{align}
The other SUSY parameters depend strongly on how the DM relic density
constraints are fulfilled and hence on the annihilation processes of
the lightest neutralino in the early universe.  In fact, a bino-like
neutralino is typically overproduced in thermal processes so that an
efficient annihilation mechanism is required in order to reproduce the
observed relic density.  There are mainly the following three
categories for the annihilation mechanism that select different regions
of the parameter space: (i) $s$-channel Higgs mediation, (ii)
co-annihilation with a light sfermion, (iii) $t$-channel sfermion
mediation.
\paragraph{Higgs-mediated annihilation.}
Case (i), in which the neutralino pair-annihilation is mediated by
Higgs bosons (mainly the CP-odd one, as the s-wave $\nt_1 \nt_1$
initial state is CP-odd), is an attractive possibility, because the
parameter space selected by this scenario unavoidably corresponds to a
large scattering cross-section with nuclei (up to $\sigma_{\rm
  SI}\simeq 10^{-41}$ cm$^2$), so that the signals at
DAMA~\cite{Bernabei:2008yi,Bernabei:2010mq},
CoGeNT~\cite{Aalseth:2010vx}, CRESST-II~\cite{Angloher:2011uu} and the
three events recently reported by CDMS~\cite{Agnese:2013rvf} could be
nicely
explained~\cite{Bottino:2002ry,Bottino:2004qi,Bottino:2008mf,Fornengo:2010mk}.
However, recent LHC results exclude light DM in this parameter region of the MSSM. 
The reason is the following: $m_{\widetilde{\chi}^{0}_{1}} \approx
10\div 20$ GeV requires a light CP-odd Higgs boson $A$ (with $m_{A}
\simeq 100$ GeV) and quite large values of $\tan \beta$ ($\gtrsim 35$)
to make the annihilation cross-section efficient
enough~\cite{Kuflik:2010ah,Feldman:2010ke,Vasquez:2010ru,Calibbi:2011ug,Calibbi:2011un}.
This setup has been recently excluded by extra Higgs boson searches at
the LHC~\cite{Chatrchyan:2012vp,Aad:2012dn}. On the other hand, larger
values of $m_{\widetilde{\chi}^{0}_{1}}$ ($20\div 30$ GeV) still
correspond to a direct detection cross-section $\sigma_{\rm SI}\simeq
10^{-42}\div 10^{-41}$ cm$^2$ that is excluded by
XENON100~\cite{Aprile:2012nq} for that mass
range~\cite{Calibbi:2011un}.\footnote{In addition, the recent LHCb
  evidence for the rare decay $B_{s}\rightarrow \mu^{+} \mu^{-}$ with
  BR $\approx 3\times 10^{-9}$~\cite{Aaij:2012ct} and the observation
  of a SM-like Higgs with $m_h\approx 125$
  GeV~\cite{aad:2012gk,chatrchyan:2012gu} are also incompatible with
  the parameter space of the Higgs mediation scenario, as one can see
  in Ref.~\cite{Calibbi:2011ug}.}
In other words, LHC searches for the CP-odd Higgs in combination with
direct dark matter searches constrain $m_A$ to values that cannot
efficiently mediate neutralino annihilation for
$m_{\widetilde{\chi}^{0}_{1}} \lesssim 30$ GeV.
\paragraph{Co-annihilation with sfermions.}
The co-annihilation scenario (ii) with a light stau has recently
awaken a great interest, because the parameter choice can be also
compatible with an enhancement of the $h\to \gamma \gamma$ decay rate,
as discussed in Ref.~\cite{Carena:2012gp}.  However, in order to have an
efficient co-annihilation for a neutralino lighter than 30 GeV, one
would need $m_{\stau}\lesssim 80$ GeV \cite{Carena:2012gp}, below the
limit set by the LEP experiments \cite{PDG}. On the other hand, if the
stau-neutralino mass splitting is small enough to evade direct LEP
searches, the scenario is seriously challenged by $Z$ width
measurements implying $m_{\stau}\gtrsim 40$ GeV~\cite{PDG} (unless the
stau left-right mixing is tuned such that the coupling to $Z$ is strongly
suppressed) and would anyway correspond to a too efficient
annihilation, i.e.~to a DM relic density way below the observed value.
As we will discuss in the following, neutralino DM with
$m_{\widetilde{\chi}^{0}_{1}} \lesssim 30$ GeV also requires a light
stau. However, the selected parameter space requires small values of
the Higgsino mass parameter $\mu$, while very large values of the stau
left-right mixing $\mu\tan\beta$ are needed in order to have an effective
enhancement of $\Gamma(h\rightarrow \gamma \gamma)$
\cite{Carena:2011aa,Carena:2012gp}.  We have checked that the
parameter region we consider does not drive the co-annihilation
process and does not give a significant enhancement of
$\Gamma(h\rightarrow \gamma \gamma)$.  Nevertheless, the current LHC
measurements of the Higgs decay rates, especially $\Gamma(h
\rightarrow {\rm invisible})$ still provide important information for
our scenario.  We will come back to this point later.

\paragraph{Sfermion mediation.}
The only option left is $t$-channel sfermion mediation (iii).  Light
neutralino DM is indeed possible in presence of a light stau, as
recently shown in Refs.~\cite{AlbornozVasquez:2011yq,Grothaus:2012js}.  The
reasons why the exchange of other sfermions cannot give enough
enhancement to the annihilation cross-section are twofold: (a) LEP and
LHC searches put severe limits on the masses of the other sfermions,
(b) efficient annihilation of very light neutralinos requires the
contribution of Yukawa interactions.

The annihilation cross-section is inverse-proportional to the mass of
the mediating sfermion. Therefore, it is constrained by the bound on
the mass of the mediation field, the so-called Lee-Weinberg
bound~\cite{Lee:1977ua}.  The LHC places strong limits on the masses
of squarks and first and second generation sleptons. For instance,
direct slepton searches with $e$ and $\mu$ in final states at the LHC
imply $m_{\widetilde \ell}\gtrsim 275$ GeV
(${\widetilde \ell}={\widetilde e},{\widetilde\mu}$) for
$m_{\nt_1}\lesssim 30$ GeV~\cite{CMS:aro}.  Similarly, the bounds on
direct production of stop and sbottom subsequently decaying to the
lightest SUSY particle (LSP) are $m_{\widetilde t},~ m_{\widetilde
  b}\gtrsim 650$ GeV~\cite{ATLAS:2013cma}.  If the sfermion-neutralino
mass-splitting is small enough ($\lesssim 10$ GeV), these collider
bounds can be evaded, as the outgoing fermion is too soft to be
detected (see e.g.~\cite{Arbey:2012na}). However, LEP observations of
the $Z$ decays imply the lower bound on any sfermion $m_{\widetilde
  f}\gtrsim 40$ GeV. This means that for $m_{\nt_1}\lesssim 30$ GeV it
is anyway not possible to have efficient annihilation mediated by
light sfermions and to evade at the same time the collider
constraints. This justifies {\it a posteriori} our choice of
concentrating on a neutralino lighter than 30 GeV.\footnote{In
  principle, one could have a tuned scenario with few GeV of
  neutralino-sbottom mass splitting and left-right sbottom mixing such
  that the $Z\tilde{b}\tilde{b}$ interaction gets strongly
  suppressed~\cite{Arbey:2012na}. This possibility might be
  challenged by indirect DM searches through antiproton and gamma-ray
  production~\cite{Asano:2011ik}. See however \cite{Arbey:2013aba}.}
\begin{figure}[t]
\centering
\includegraphics[width=1.\textwidth]{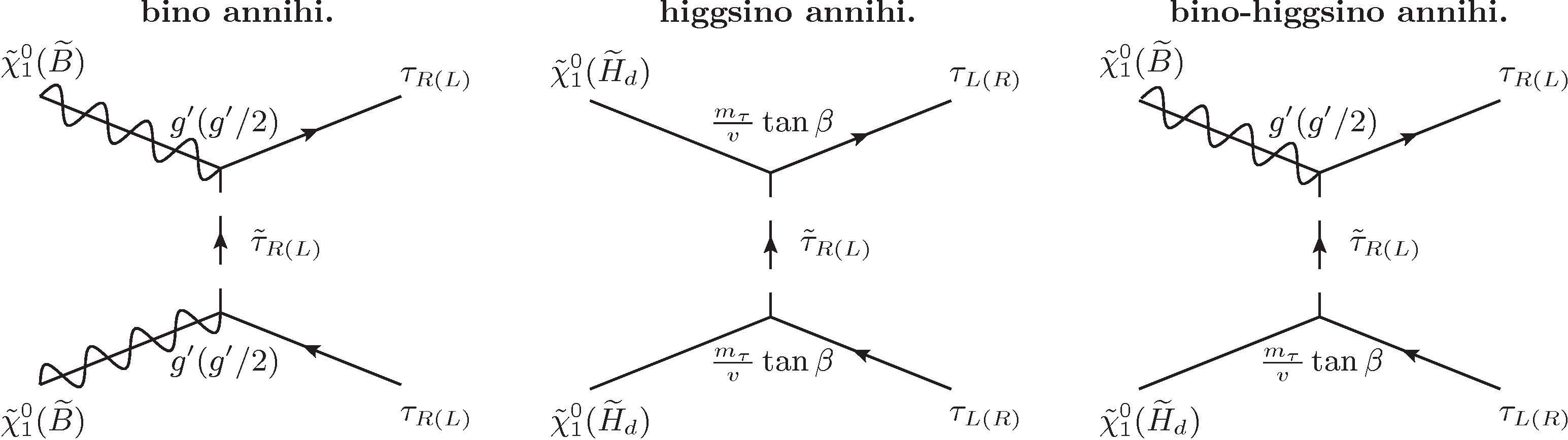}
\caption{\label{Fig:annihilation}
Relevant neutralino annihilation processes 
mediated by a light stau.}
\end{figure}

The only option left is that the lightest neutralino, that is mainly
bino, does pair-annihilate into $\tau^+\tau^-$ via the $t$-channel stau
exchange diagram driven by gauge interactions shown in the left panel
of Fig.~\ref{Fig:annihilation}.  Since the amplitude is proportional
to the square of the hypercharge of the mediating sfermion,
the right-handed stau mediation is much more efficient than the
left-handed one, giving a factor 16 larger cross-section.
As a consequence, the relic density constraints select the mass of the
right-handed stau to be not far above the LEP lower
bound~\cite{AlbornozVasquez:2011yq}:
\begin{align}
m_{\widetilde{\tau}_{R}} \sim 100~\text{GeV}.
\label{eq:Our-parameter-region-Nr2}
\end{align}

However, the gauge interaction diagram might not be enough to have an
efficient neutralino annihilation, especially for even lower
neutralino masses, e.g.~$\lesssim 20$ GeV.
If the lightest neutralino has a sizeable Higgsino component, this can
contribute through Higgsino-bino and Higgsino-pair annihilation
diagrams (shown as the middle and the right diagrams in
Fig.~\ref{Fig:annihilation}).  These further contributions assist in
realizing the correct relic density with a very light neutralino, in
particular for large values of $\tan \beta$.\footnote{Instead, a light
  wino does not increase the annihilation cross-section enough, if
  Higgsinos are not light too. The reason is mainly that the wino
  component in $\nt_1$ vanishes if Higgsinos decouple.}  A significant
Higgsino component in the lightest neutralino requires a small
$\mu$-term. This implies that, in addition to the lightest $\stau$,
the lightest chargino $\ch_1$ and two Higgsino-like neutralinos
($\nt_2$ and $\nt_3$, if $M_2 \gg \mu$) are necessarily light.

In short, in order to enhance the annihilation process driven by the
Higgsino component in the lightest neutralino, the following values of
the above parameters are selected:
\begin{align}
m_{\nt_1}\lesssim 30\,(20) \,\,{\rm GeV} \qquad\Rightarrow\qquad
\mu \sim 100 \text{ GeV}, \qquad \tan \beta \gtrsim 10\,(30)~.
\label{eq:Our-parameter-region-Nr3}
\end{align}

In summary, the parameter space compatible with light neutralino DM we
aim at studying is essentially determined by the values of the four
parameters given in Eqs.~(\ref{eq:Our-parameter-region-Nr1}-\ref{eq:Our-parameter-region-Nr3}), ($M_{1}$, $\mu$,
$m_{\widetilde{\tau}_{R}}$, and $\tan \beta$) and does not depend on
the detail of the other SUSY parameters.  In the following we 
illustrate numerically the parameter space qualitatively
sketched above. We then make use of the simplified model defined by
this subset of the parameters in order to discuss the limits set by
the LHC on neutralino DM.

\section{Scan of the parameter space and experimental constraints}
\label{sec:parspace}	
In order to investigate the parameter space selected by requiring
light neutralino dark matter, we perform a numerical scan by means of
the {\tt SuSpect} \cite{Djouadi:2002ze} and {\tt micrOMEGAs}
codes~\cite{micro}.  The low-energy values of the four parameters
identified above were randomly varied in the following ranges:
\begin{eqnarray}
 10~{\rm GeV} \le ~M_1~ \le  45~{\rm GeV},  \quad
 65~{\rm GeV} \le ~m_{\stau_R}~ \le  200~{\rm GeV},\nonumber \\
 90~{\rm GeV} \le ~\mu~ \le  400~{\rm GeV},  \quad 5 \le ~\tan\beta~
 \le  60 \,,
\end{eqnarray}
and the SUSY breaking scale within {\tt SuSpect} is set to the approximate scale 
of the relevant low-mass states ($100$ GeV). The other SUSY parameters have 
a marginal role in fulfilling the relic density constraints and they were set 
to the following constant values:
\begin{align}
m_{\widetilde f} = M_3 = m_A = 2~{\rm TeV},~ M_2 = 1~{\rm TeV},~ A_t =
1.5\times m_{\widetilde f} \,,
\end{align}
where $m_{\widetilde f}$ denotes all sfermion soft masses besides
$m_{\stau_R}$, $M_i$ are the gaugino masses, $m_A$ the CP-odd Higgs
mass. The values chosen for $A_t$ and $m_{\widetilde f}$ give the
Higgs mass consistent with the LHC measurements, $m_h \approx 125$ GeV. 
All other $A$-terms are set to zero.

In the following we describe various (possible) constraints on the
resulting parameter space, where, as motivated above, we limit our
discussion on the light neutralino regime defined in
Eq.~(\ref{eq:mchi30}).

\subsection{Relic density, LEP and other standard constraints}
\label{sec:relicdensity}
\paragraph{DM relic density.}
Assuming a standard thermal history of the universe, we compute with
{\tt micrOMEGAs}~\cite{micro} the neutralino relic density and impose
a conservative 3$\sigma$ upper bound taken from Ref.~\cite{Ade:2013zuv},
\begin{equation}
\Omega_{\rm DM}h^2 \le 0.124 \,.
\label{eq:planck}
\end{equation}
\paragraph{Direct SUSY searches at LEP.} 
The 95\% CL LEP bounds on the lightest stau and chargino masses listed
in Ref.~\cite{PDG} are, respectively,
\begin{equation}
\label{eq:LEPlimits}
m_{\stau_R}\ge 81.9~\text{GeV}\,,\quad \text{and} \qquad
m_{\ch_1} \ge 94~\text{GeV}.
\end{equation} 
In addition, we consider bounds from searches of $\nt_1 \nt_{2,3}$
associated production at LEP, followed by the decay $\nt_{2,3} \to
\nt_1 Z^{(*)}$. The conservative limit on this process is about 100
$\rm fb$ for the $\nt_1$ mass range we are interested 
in~\cite{Abbiendi:2003sc}. Explicitly this bound reads
\begin{equation}
\sum_{k=2,3} \sigma (e^+ e^- \to \nt_1 \nt_k) \times {\rm BR}(\nt_k \to \nt_1 Z^{(*)}) < 100~{\rm fb}.  
\label{eq:opal}
\end{equation}
We calculate the associated production cross-sections at LEP using the
leading order formulae reported in Refs.~\cite{Ellis:1983er,Bartl:1986hp},
the branching fractions were computed by means of the {\tt SUSY-HIT}
package~\cite{Djouadi:2006bz}.
\paragraph{$Z$ invisible width.} 
\begin{figure}[t]
\centering
 \includegraphics[width=0.6\textwidth]{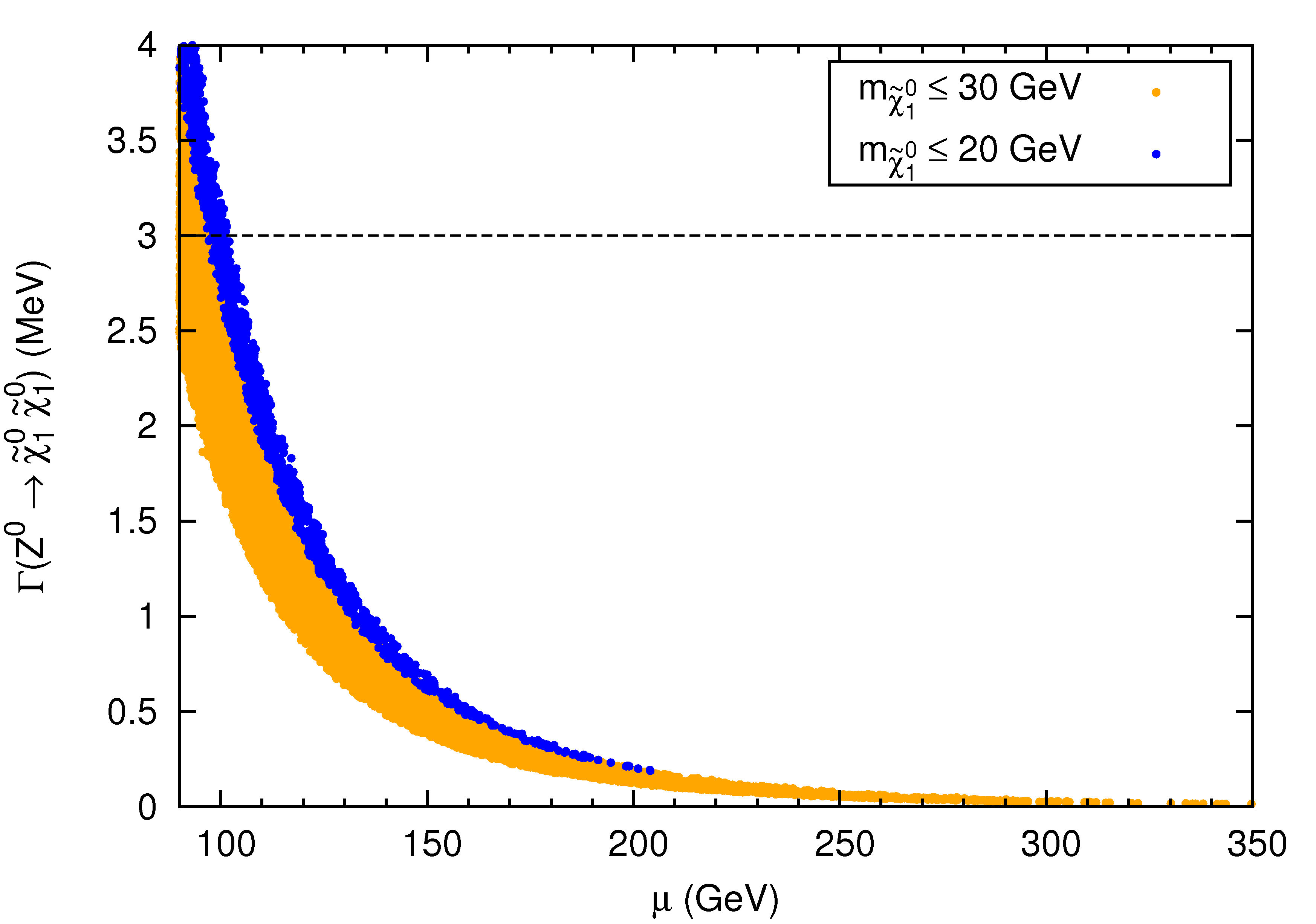}
 \caption{Contribution to the invisible $Z$ width form the process
   $Z\to \nt_1 \nt_1$ as a function of the Higgsino mass parameter
   $\mu$. The dashed line corresponds to the LEP limit of
   Ref.~\cite{ALEPH:2005ab}.\label{fig:Zchichi}}
\end{figure}
As discussed above, a sizeable Higgsino $\widetilde{H}_d$ component in
$\nt_1$ is required. This gives rise to the coupling $\nt_1 \nt_1 Z$
and also to $\nt_1 \nt_1 h$ that originates from the gauge vertex
$\widetilde{B} H_d \widetilde{H}_d$.  As a consequence, the decays
$Z\to \nt_1\nt_1$ and $h\to \nt_1\nt_1$ can occur at relevant rates
and we expect stringent constraints from the LEP measurement of the
invisible $Z$ width as well as from Higgs observations at the LHC, as
recently discussed in Ref.~\cite{Dreiner:2012ex}.  The decay width of
the $Z$ boson into a lightest neutralino pair
reads~\cite{Barbieri:1987hb}:
\begin{equation}
  \label{eq:Zinv}
    \Gamma (Z \rightarrow \widetilde{\chi}^{0}_{1} \widetilde{\chi}^{0}_{1}) =
  \frac{G_F}{\sqrt{2}} \frac{M_Z^3}{12\pi}  \left( 1 - \frac{4
      m_{\widetilde{\chi}^{0}_{1}}^{2}}{M_{Z}^{2}} \right)^{\frac{3}{2}} \left|
    N_{13}^{2} - N_{14}^{2} \right|^{2} \,,
\end{equation}
where $N$ is the neutralino mixing matrix defined by 
$\nt_i = N_{i1} \widetilde{B} + N_{i2} \widetilde{W} + 
N_{i3} \widetilde{H}_d + N_{i4} \widetilde{H}_u$.

In Fig.~\ref{fig:Zchichi} we plot the resulting $\Gamma(Z\to
\nt_1\nt_1)$ as a function of $\mu$, after applying the other
constraints described above. The orange points correspond to $\nt_1$
DM with $m_{\nt_1} \leq 30$ GeV, the blue points to $m_{\nt_1} \leq
20$ GeV.  The figure is to be compared to the LEP bound on the new
physics contribution to $\Gamma(Z \to {\rm invisible})$, $\Delta
\Gamma_Z^{\rm inv}$~\cite{ALEPH:2005ab}:
\begin{align}
 \Delta \Gamma_Z^{\rm inv} < 3  ~{\rm MeV}\quad(95\%~{\rm CL}).
\end{align}
As we see, only a small number of points (corresponding to very low
$\mu$ and particularly light $\nt_1$ masses) is excluded by this
observable.\footnote{We verified that these results are hardly affected
  by NLO corrections, as presented in Ref.~\cite{Heinemeyer:2007bw}, where 
  a factor of $1/2$ is missing in the tree-level
  result stated in Eq.~(51).}  A similar
constraint from $h\to \nt_1\nt_1$ will be discussed in section
\ref{sec:invH}.
\paragraph{Flavor processes.} 
Rare decays such as $B_{s}\rightarrow \mu^{+} \mu^{-}$, recently
measured at $\approx 3\sigma$ by LHCb \cite{Aaij:2012ct}, the
partially correlated processes of the kind $b\to s\gamma$, charged
current processes like $B \rightarrow \tau \nu$ and $K \rightarrow \mu
\nu$ set stringent constraints on supersymmetric models with large
$\tan\beta$ \cite{Buras:2002vd,Isidori:2006pk} (for a discussion in the context of light
neutralino DM, see~\cite{Calibbi:2011ug}). Nevertheless, these
processes are mediated by the charged or CP-odd Higgs boson 
(and also squarks in the case of $b\to s\gamma$), hence the rates
depend on the mass-scale of the extended Higgs sector (and squarks).
These parameters can be set to arbitrarily high values, as the $\nt_1$
annihilation cross-section is not affected by them. As a consequence,
in general flavor processes do not constrain the light neutralino
parameter space we consider here.
\paragraph{LHC searches.} Recent limits from chargino searches in
multi-leptons + missing transverse energy 
$\slashed{E}_T$ events at the
LHC~\cite{ATLAS:2013rla,CMS:aro} depend on the mass of the first and
second generation sleptons and assume wino-like charginos.  Hence they
do not apply to our scenario. Bounds from LHC searches based on at least
one pair of hadronically decaying taus~\cite{ATLAS:2013yla} are not
applied to the scan we are presenting. They will be discussed in
detail in the next sections.
\begin{figure}[t]
\centering
\includegraphics[width=0.6\textwidth]{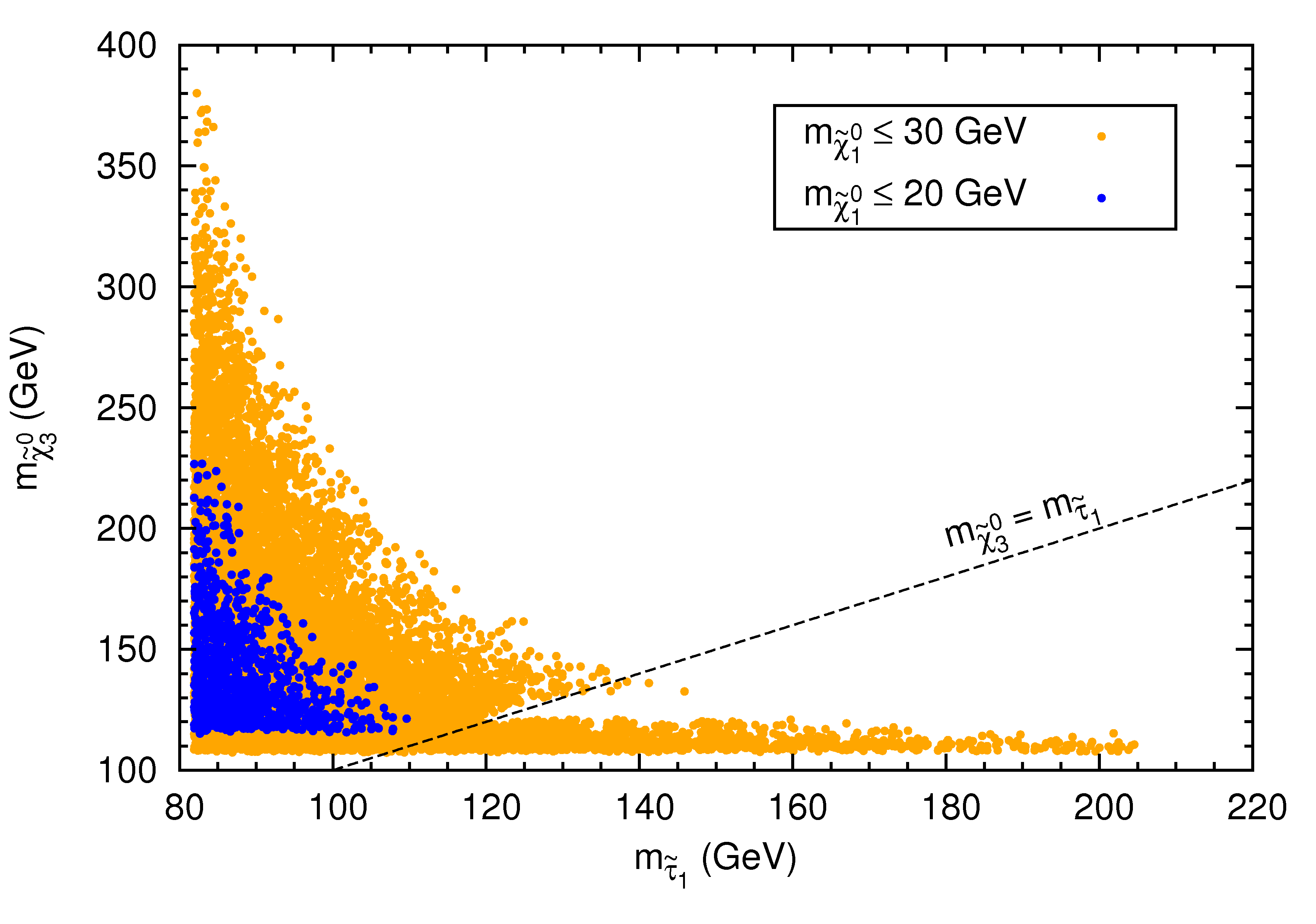}
\caption{Viable parameter space after applying the constraints
  discussed in the text, displayed on the $m_{\stau_1} - m_{\nt_3}$
  plane.\label{fig:stau-chi}}
\end{figure}
\paragraph{} 
The resulting parameter space compatible with light neutralino dark
matter and the bounds listed above is shown in
Fig.~\ref{fig:stau-chi}, for the physical masses $m_{\stau_1}$ and
$m_{\nt_3}$. Again the orange points correspond to $m_{\nt_1} \leq 30$
GeV, the blue points to $m_{\nt_1} \leq 20$ GeV. The upper boundary of
the relevant parameter space is set by the CMB constraint,
Eq.~(\ref{eq:planck}), that requires low values for either
$m_{\stau_1}$ or $\mu\approx m_{\nt_3}$.  Hence, light neutralino dark
matter implies upper bounds on the stau and Higgsino masses:
\begin{equation}
  m_{\stau_1}\lesssim 210~{\rm GeV}, \quad m_{\ch_1} \approx m_{\nt_2}
  \approx m_{\nt_3} 
\lesssim 380 ~{\rm GeV}.
\label{eq:up-bounds}
\end{equation}
Notice that, below the threshold where $\nt_{2,3}$ decays into a real
stau are possible, some points are excluded by the limit from LEP
neutralino searches stated in Eq.~(\ref{eq:opal}).  This is a
consequence of the fact that, as far as $\nt_{2,3}$ decays into a real
$Z$ are kinematically allowed while the decays into a real stau are
forbidden, one obtains ${\rm BR}(\nt_{2,3} \to \nt_1 Z)\approx 1$, so
that the LEP bound is maximized. On the other hand, for very low
$m_{\nt_{2,3}}$, where this decay is not open, or where this decay
competes with on-shell decays into staus, constraints from
Eq.~(\ref{eq:opal}) vanish.
\begin{figure}[t]
\centering
\includegraphics[width=0.6\textwidth]{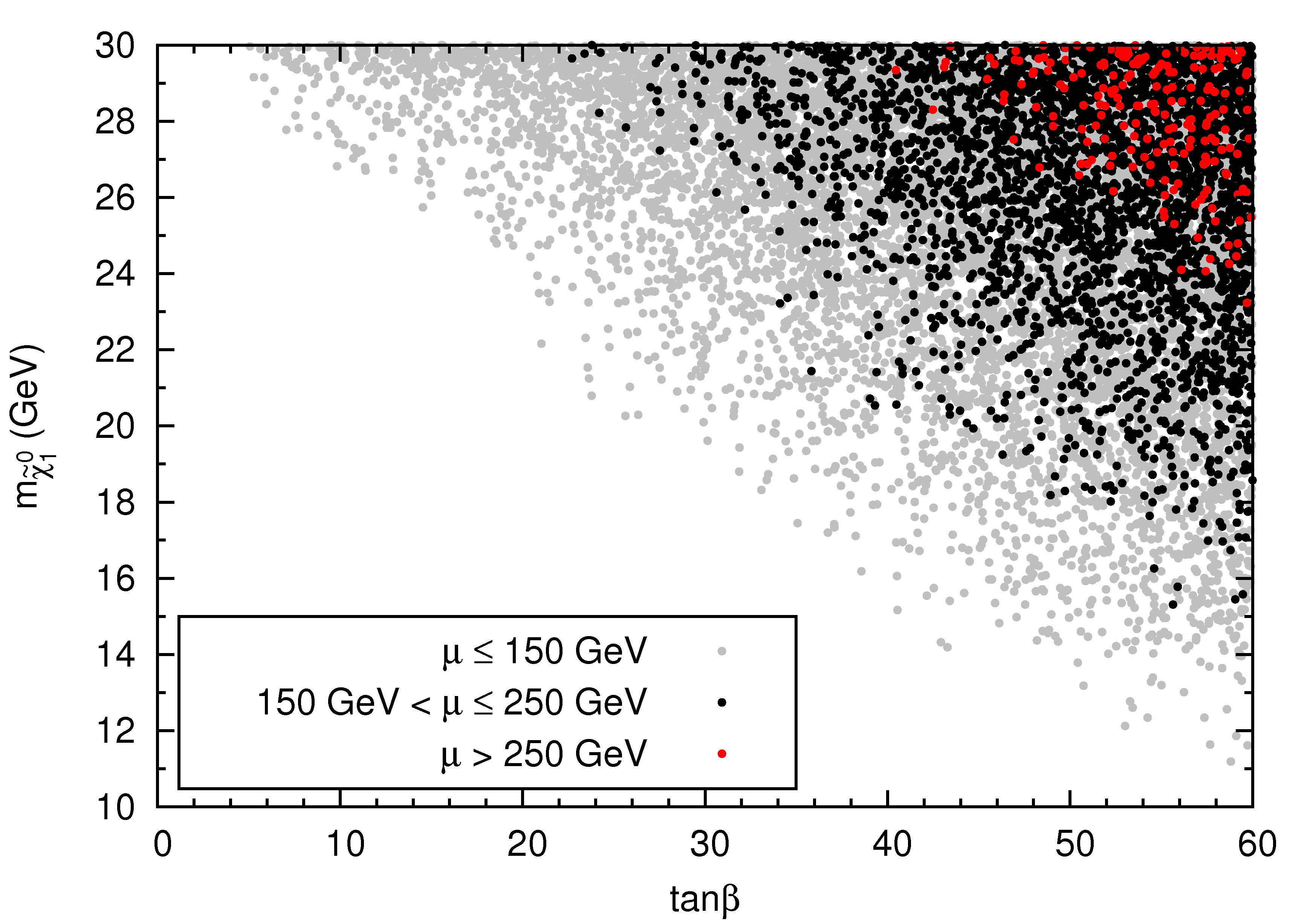}
\caption{Viable parameter space after applying the constraints
  discussed in the text, displayed on the $\tan\beta - m_{\nt_1}$
  plane.\label{fig:tb-chi}}
\end{figure}

In Fig.~\ref{fig:tb-chi}, the same parameter space is shown in the
$\tan\beta - m_{\nt_1}$ plane. Different ranges for $\mu$ are
displayed in different colors. As a consequence of the Higgsino
contribution to the $\nt_1$ annihilation, the larger $\tan\beta$ is,
the smaller $m_{\nt_1}$ can be.  Taking $\tan\beta < 60$ GeV, required
for perturbativity of the bottom Yukawa coupling, one then gets the
following lower bound on the mass of neutralino dark matter:
\begin{equation}
 m_{\nt_1} \gtrsim 11 ~{\rm GeV}.
\end{equation}
{}From the figure we see that neutralino masses close to this lower
limit require a particularly light Higgsino sector, namely $\mu
\lesssim 150$ GeV.

In sections \ref{sec:LHC_pheno}-\ref{sec:LHC_numerics}, we are
going to discuss the consequences for SUSY searches at the LHC of the
spectrum discussed above, including all constraints listed here.
But first, let us briefly discuss the possible impact of direct and
indirect DM searches on the light neutralino parameter space.

\subsection{Direct and indirect Dark Matter searches}
\label{sec:DMsearches}
Several works have been recently dedicated to light neutralinos (see
e.g.~\cite{Kuflik:2010ah,Feldman:2010ke,Vasquez:2010ru,Fornengo:2010mk,Calibbi:2011ug,Arbey:2012na,Boehm:2013qva}),
aiming at a possible explanation of the annual modulation signal at
DAMA/LIBRA~\cite{Bernabei:2008yi,Bernabei:2010mq} and also reported by
CoGeNT~\cite{Aalseth:2010vx} and the excess of nuclear recoil events
observed by CoGeNT itself and CRESST~\cite{Angloher:2011uu}.
Recently, three signal events have been also reported by
CDMS~\cite{Agnese:2013rvf}. Although the situation is not conclusive
yet (the above claims are not consistent with the XENON100
results~\cite{Aprile:2012nq} and do not even seem to be perfectly
compatible with each other
\cite{Kopp:2009qt,Schwetz:2011xm,Frandsen:2013cna}), the above results
might be broadly accounted for by a light WIMP with a mass in the
range of $\ord{10}$ GeV and a spin-independent (SI) elastic DM-nucleon
cross-section, $\sigma_{\rm SI}$, of the order of $10^{-42}$ to
$10^{-40}$~cm$^2$.  As previously mentioned, it is not possible to
realize the above configuration within the
MSSM~\cite{Calibbi:2011ug,Calibbi:2011un,Vasquez:2010ru}: in fact,
such a relatively large spin-independent cross-sections require a
light extended Higgs sector, which is now excluded for moderate to
large $\tan\beta$ by (i) searches at the LHC for extra Higgs bosons
decaying into pairs of taus~\cite{Chatrchyan:2012vp,Aad:2012dn}, and
(ii) the recent observation of the decay $B_{s}\rightarrow \mu^{+}
\mu^{-}$ at LHCb~\cite{Aaij:2012ct} with a branching fraction
compatible with the SM prediction \cite{Buras:2012ru}.

\begin{figure}[t]
\centering
 \includegraphics[width=0.45\textwidth]{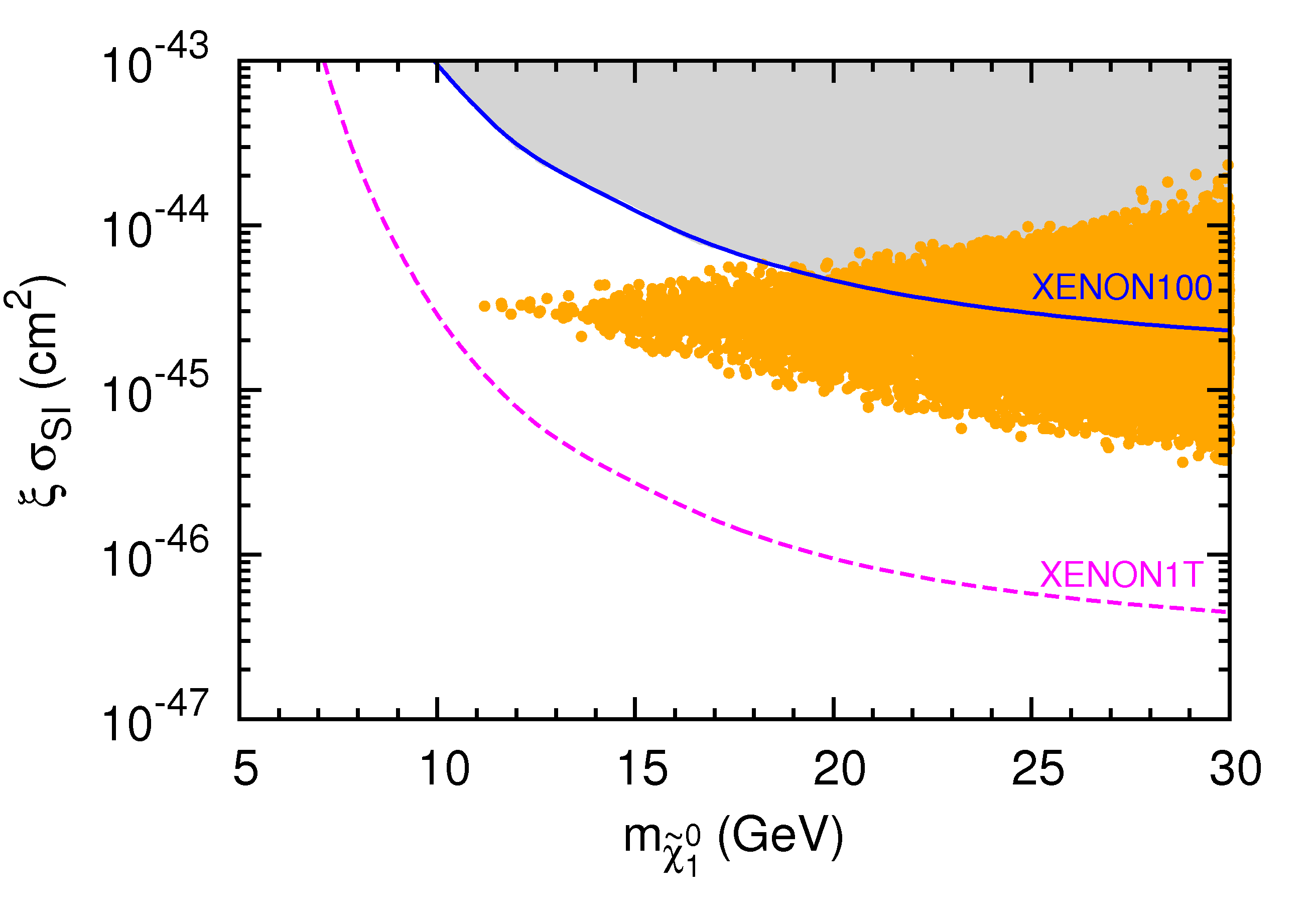}
 \includegraphics[width=0.45\textwidth]{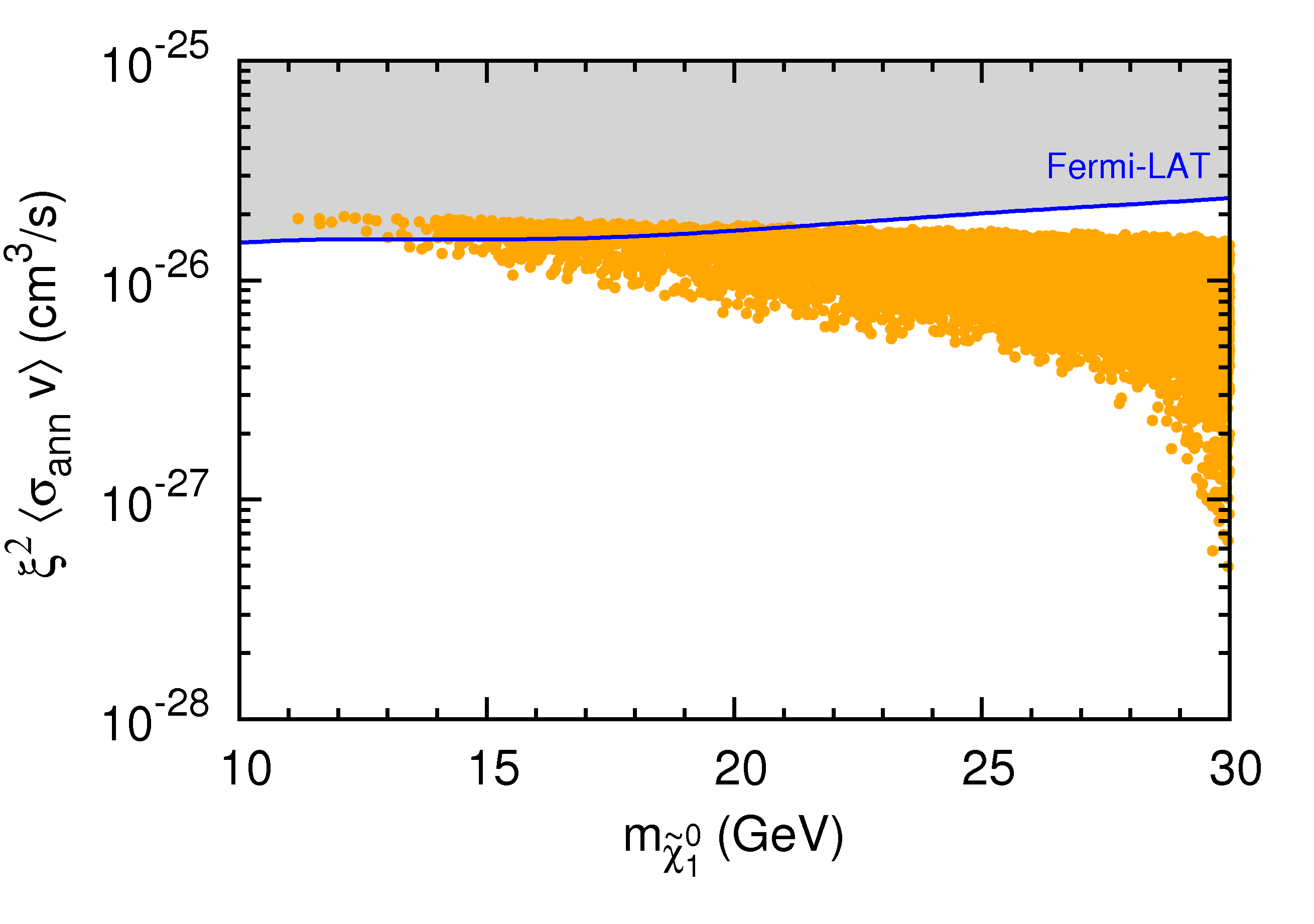}
 \caption{Model predictions and experimental limits on the SI
   scattering cross-section with nuclei (left panel) and the velocity-averaged annihilation
   cross-section (right panel) as a function of $m_{\nt_1}$. 
The rescaling factor $\xi\equiv \Omega_{\rm DM}h^2/0.119$ is taken into account.
   See the text for the details.\label{fig:DMsearches} }
\end{figure}

Direct detection experiments might still set relevant constraints on
the region we are exploring, as there is an irreducible contribution
to $\sigma_{\rm SI}$ mediated by the exchange of the light Higgs $h$,
whose coupling with the lightest neutralino is set by the relic
density bound. Other contributions, such as the squark-mediated ones,
are model-dependent as they are controlled by parameters that are not
constrained by $\Omega_{\rm DM}$.  The calculation of the scattering
cross-section is still affected by a residual model dependence, coming
from the fact that the decoupling of the contribution mediated by the
heavy CP-even Higgs $H$ is very slow: in fact both $h$ and $H$
contributions are enhanced by a small value of $\mu$ (that implies a
large Higgsino component in $\nt_1$), though solely the $H$
contribution is enhanced by large $\tan\beta$. Therefore, if one
keeps, as in our scan, $m_H \approx m_A \approx \ord{1}$ TeV (which
might be a reasonable choice in the light of our SUSY spectrum) and
large values of $\tan\beta$ (as required by the relic density
constraint), the size of the two contributions is comparable, and the
SI cross-section can be a powerful probe of the corner of the
parameter space with small values of $\mu$.
Indeed, such a region is difficult to directly probe at the LHC, 
as we will see in section \ref{sec:LHC_numerics}.
As an illustration, in the left panel of Fig.\ref{fig:DMsearches} we plot
$\sigma_{\rm SI}$ (weighted by $\xi\equiv \Omega_{\rm DM}h^2/0.119$)
versus $m_{\nt_1}$ for $m_A = 2$ TeV and the default
{\tt micrOMEGAs} value of the quark masses and the hadronic matrix
elements.\footnote{A variation of these parameters in the ranges
  reported in Ref.~\cite{micro} can lower the prediction for $\sigma_{\rm SI}$ 
  by about a factor of two.}  As we can see, XENON100 already
represents a relevant constraint on the light neutralino parameter
space and the expected sensitivity of XENON1T has the potential of
completely testing our scenario.  We checked, however, that, if we allow
$m_A$ to be $\ord{10}$ TeV, the $H$ contribution would finally
decouple and the SI cross-section would be decreased by almost a
factor of three, so that most of the points would escape from the XENON100
bound, although probably still being within the future XENON1T sensitivity.

Contrary to $\sigma_{\rm SI}$, the prediction for the spin-dependent
(SD) cross-section, $\sigma_{\rm SD}$, is much more robust, as the
relevant dimension-6 operator $\bar{\tilde{\chi}}^0_1 \gamma^\mu
\gamma_5 \nt_1~\bar{q}\gamma_\mu \gamma_5 q$ is in our case mediated
by a $Z$ exchange, which at leading order does not depend on the SUSY parameters except
for the neutralino mixing parameters. Bounds on the $\sigma_{\rm SD}$ from direct detection
experiments such as PICASSO~\cite{Archambault:2012pm},
COUPP~\cite{Behnke:2012ys}, SIMPLE~\cite{Felizardo:2011uw},
KIMS~\cite{Kim:2012rza} IceCube~\cite{Aartsen:2012ef},
XENON100~\cite{Aprile:2013doa} (see also \cite{Garny:2012it}) and 
the Baksan Neutrino Observatory \cite{Boliev:2013ai}
are not much constraining yet for light neutralino DM. Nevertheless, LHC
searches for events with a monojet plus missing transverse
momentum~\cite{ATLAS:2012ky,Chatrchyan:2012me} can be translated into
bounds on neutralino-nucleon scattering cross-sections
\cite{Birkedal:2004xn,Beltran:2010ww,Rajaraman:2011wf,Fox:2011pm}.  In
the case of the operator given above, the bounds on $\sigma_{\rm SD}$
translated from the monojet searches are much stronger than the ones
obtained from underground direct DM detection experiments.  As we are
going to see in section~\ref{sec:monojet}, searches for monojet events
performed by CMS and ATLAS might set a strong constraint on the
small-$\mu$ region of the parameter space.

Let us now briefly discuss indirect detection limits.  Our scenario
predicts as a unique annihilation channel $\nt_1\nt_1 \to
\tau^+\tau^-$.  The most stringent bound on this channel is provided
by the observation of gamma-rays from satellite galaxies performed by
the Fermi-LAT collaboration~\cite{Ackermann:2011wa}.  The limit is
shown in the right panel of Fig.~\ref{fig:DMsearches} and it is
compared with our parameter space prediction for the velocity-averaged
annihilation cross-section (weighted by $\xi^2$) as
computed with {\tt micrOMEGAs}.  As we can see, the Fermi-LAT limit
slightly increases the lower bound of the neutralino mass to about 13
GeV.  Still, a conservative estimate of about $30\%$ of possible
uncertainties affecting the theoretical computation and the
experimental limit could easily make such a bound
milder. Nevertheless, the model predictions lie certainly on the
border of the current Fermi-LAT sensitivity so that the light
neutralino scenario might be tested independently in the near future
by gamma-ray observations.

A detailed discussion of experimental and theoretical uncertainties
affecting direct and indirect DM searches is beyond the scope of this
paper. What we can conclude from this brief discussion is that direct
and indirect DM searches are approaching the sensitivity for testing
the light neutralino scenario, so that we can expect this to be
achieved in the upcoming years, but at the moment we can
conservatively consider most of our parameter space unconstrained by
such searches.  Therefore, we are not going to impose the bounds from
XENON100 and Fermi-LAT further in this study and concentrate on the
constraints that can be obtained from the LHC data.

\section{Spectrum and LHC phenomenology}
\label{sec:LHC_pheno}

As illustrated in the previous sections, light neutralino DM in the
MSSM is only consistent with WMAP and Planck observations in a region
of the parameter space with peculiar features: a small soft mass for
the right-handed stau, a small Higgsino mass parameter $\mu$ and
moderate to large $\tan\beta$.
This leads us naturally to the following spectrum at the electroweak
scale: besides the lightest neutralino that is assumed to have mass
$m_{\nt_{1}} \lesssim 30$ GeV, there are only the lighter stau (mainly
right-handed), the Higgsino-dominated neutralinos $\nt_{2,3}$, and the
lighter chargino $\ch_1$, which take masses that are not much above
$\approx 100$ GeV.  More specifically, these states are assumed to be
in the range between the LEP lower bounds given in
Eq.~(\ref{eq:LEPlimits}) and the upper limits given in
Eq.~(\ref{eq:up-bounds}), i.e. $94~{\rm GeV} < m_{\ch_{1}} < 380~{\rm GeV}$. 
All the other states (in particular the strong-interacting
superpartners) play no role in satisfying the relic density constraint
and can be in principle too heavy to be detected by LHC experiments.
Still, direct electroweak production for $\stau_1$, $\nt_{2,3}$
and $\ch_1$ with masses of $\ord{100}$ GeV can be sizeable, with
cross-sections up to the pb level for proton-proton collisions at
$\sqrt{s}=14$ TeV~\cite{Baer:1993ew,Eckel:2011pw}.  This represents
the only unavoidable contribution to the total SUSY production in our
scenario.  In order to test light neutralino DM at the LHC, it is
therefore sufficient to consider the particle content described above
with the following electroweak Drell-Yan production modes:
\begin{align}
  pp \to \stau^+_1\, \stau^-_1 +X,\quad pp \to \nt_i\, \nt_j + X ,\quad
  pp \to \nt_i\, \ch_1 +X,\quad pp \to \chp_1\, \chm_1 +X ,
\end{align}
where $i,j=2,3$.  The decays of these particles clearly depend on the
detail of the spectrum.
%%%%%%%%%%%%%%%%%%%%%%%%%%%%%%%%%%%%%%%%%%%%%%%%%%%%%%%
\begin{figure}[t]
\centering
\includegraphics[width=0.55\textwidth]{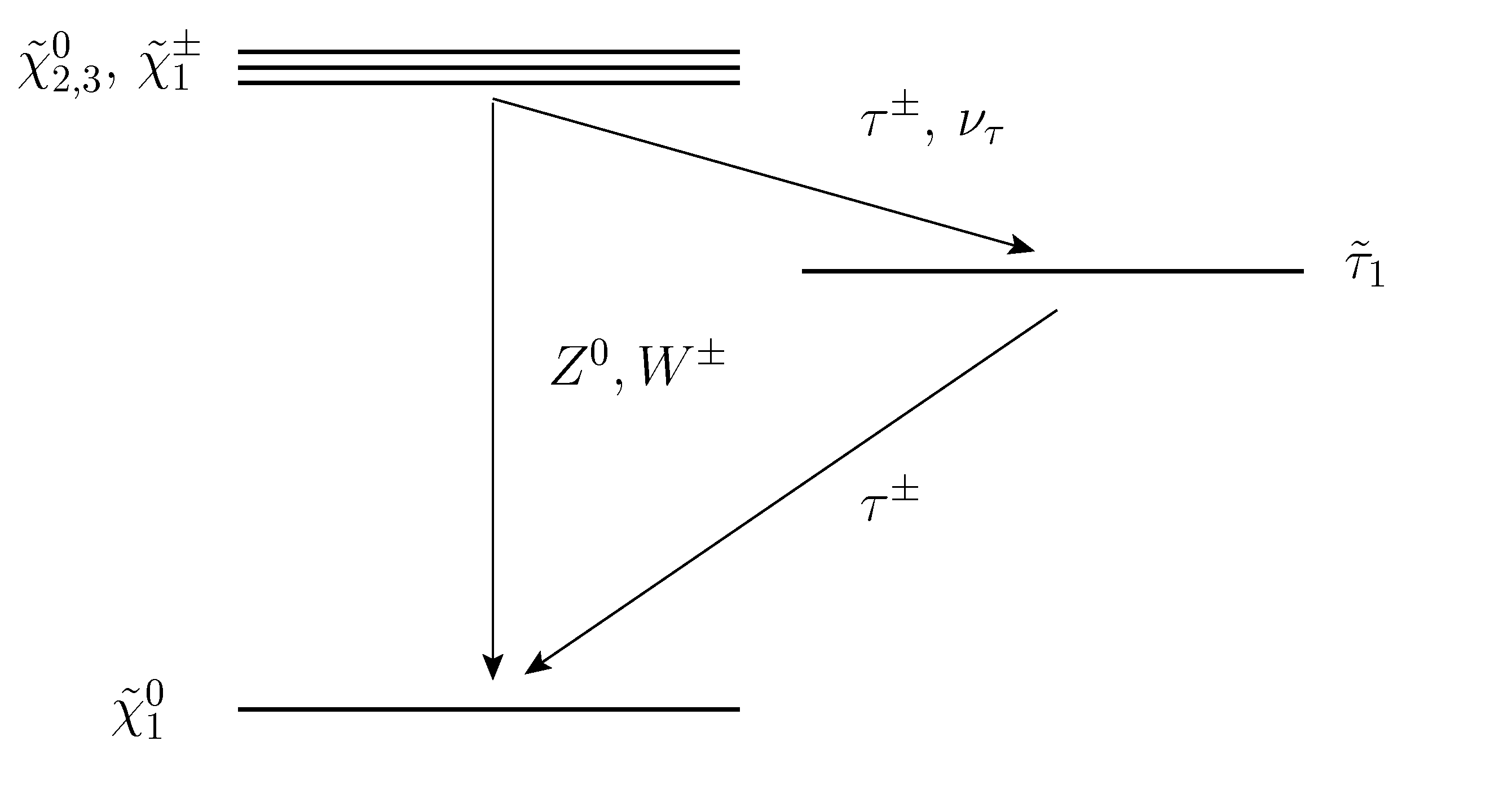}
\caption{\label{fig:spectrum} A typical mass spectrum of the light
  neutralino dark matter scenario.  The shown particles masses are
  mainly controlled by three SUSY parameters: $M_{1}$,
  $m_{\tilde{\tau}_{R}}$, and $\mu$. All the other fields are assumed
  to be heavy.}
\end{figure}
%%%%%%%%%%%%%%%%%%%%%%%%%%%%%%%%%%%%%%%%%%%%%%%%%%%%%%%

Let us first consider the hierarchy depicted in
Fig.~\ref{fig:spectrum}, i.e.,
$$m_{\ch_1} \simeq m_{\nt_{2,3}} > m_{\stau_1} > m_{\nt_1}\,,$$ 
that we typically observe in the parameter region consistent with
Planck (cf. the points above the dashed-line in
Fig.~\ref{fig:stau-chi}).  In this case the stau decays with almost
100\% probability into a tau lepton and the LSP,
\begin{align}
 \stau_1^\pm &\to \tau^\pm \nt_1 \quad [{\rm BR}\approx 100\%].
\end{align}
Charginos and heavier neutralinos can instead decay first to an
on-shell stau,
\begin{align}
 \nt_{2,3} &\to \tau^\mp \stau_1^\pm \quad [{\rm BR}\approx 90\%], \\
 \ch_1 &\to \nu_\tau \stau_1^\pm \quad [{\rm BR}\approx 75\%], 
\end{align}
where in parenthesis we show typical values for the branching
fractions for the case $m_{\ch_1} \simeq m_{\nt_{2,3}} > m_{\stau_1}$,
as computed with the {\tt SUSY-HIT} package~\cite{Djouadi:2006bz}. The
other possible channels are $Z \nt_1$ for the neutralinos and $W^\pm
\nt_1$ for the chargino. {}From the above, we see that the pair
production of Higgsino-like neutralinos can lead with high probability
($\approx$ 80\%) to a striking 4$\tau$ + $\slashed{E}_T$ signal at the
LHC from the decay chain:
\begin{align}
 \nt_{2,3}  &\to  \tau^\mp \stau_1^\pm  \to \tau^\mp \tau^\pm \nt_1.
\end{align}
Still, event rates for such a signature are suppressed by hadronic tau
reconstruction efficiencies, which are typically of the order of
$25\div 40\%$ for each hadronic tau $\tau_h$.
Similarly, $\nt_{2,3} \ch_1$ production gives about 70\% of times
events with 3$\tau$ + $\slashed{E}_T$.  Other combinations of
production and decay modes result in a smaller number of taus
(e.g.~two from $\chp_1\chm_1$ production followed by $\ch_1\to \stau_1^\pm
\nu_\tau$), hence they are, in principle, more difficult to
disentangle from the SM background.  For instance, production and
decays of SM gauge bosons and $t\bar{t}$ can copiously give signatures
like 2$\tau$ $+$ $\slashed{E}_T$. Nevertheless, the $m_{T2}$ cut, we are
going to discuss in the next section, turns out to be very efficient
in discriminating between signal and background also in this
case. Moreover, we expect a very large number of such events as
several combinations of production and decay modes contribute to this
category.  Thus, 2$\tau$ $+$ $\slashed{E}_T$ events also have an important
role in testing our scenario.

Let us now consider the case that neutralinos cannot decay to a real
stau:
$$m_{\stau_1} > m_{\nt_{2,3}} >  m_{\nt_1} \quad {\rm or} \quad 
m_{\nt_3} > m_{\stau_1} > m_{\nt_{2}} >  m_{\nt_1}. $$
Such hierarchies are consistent with the relic density bound in a
corner of the parameter space, as shown with the points below the
dashed-line in Fig.~\ref{fig:stau-chi}.  The 3-body decays $\nt_{2,3}
\to \tau^+ \tau^- \nt_1$ through an off-shell $\stau_1$ can still give
multi-tau events.  In this case, the exact branching ratios depend on
the masses of the other sfermions that can mediate 3-body decays at
comparable rates even if much heavier than $\stau_1$. We checked that
even for squarks and sleptons above the TeV scale, the branching ratio
BR($\nt_{2,3} \to \tau^+ \tau^- \nt_1$) does typically not exceed the
$15\div 20\%$ level.
Furthermore, the decay of $\nt_{2,3}$ into $\nt_1 Z$ will dominate, if
kinematically allowed. This further suppresses a possible multi-tau
signal. However, corresponding parameter regions are partly already
covered by the limit stated in Eq.~(\ref{eq:opal}). The remaining
corner of this parameter region at very small $\mu$ should be much
more difficult to directly probe at the LHC.
Nevertheless, other LHC observables already disfavor this scenario, as we
illustrate in the following.

\section[Limits from $h\to \nt_1\nt_1$ and monojet searches at the
 LHC]{Limits from $h \to \nt_1\nt_1$ and monojet searches at the LHC}
\label{sec:otherLHC}

\subsection{Invisible Higgs decays}
\label{sec:invH}
\begin{figure}[t]
\centering
 \includegraphics[width=0.45\textwidth]{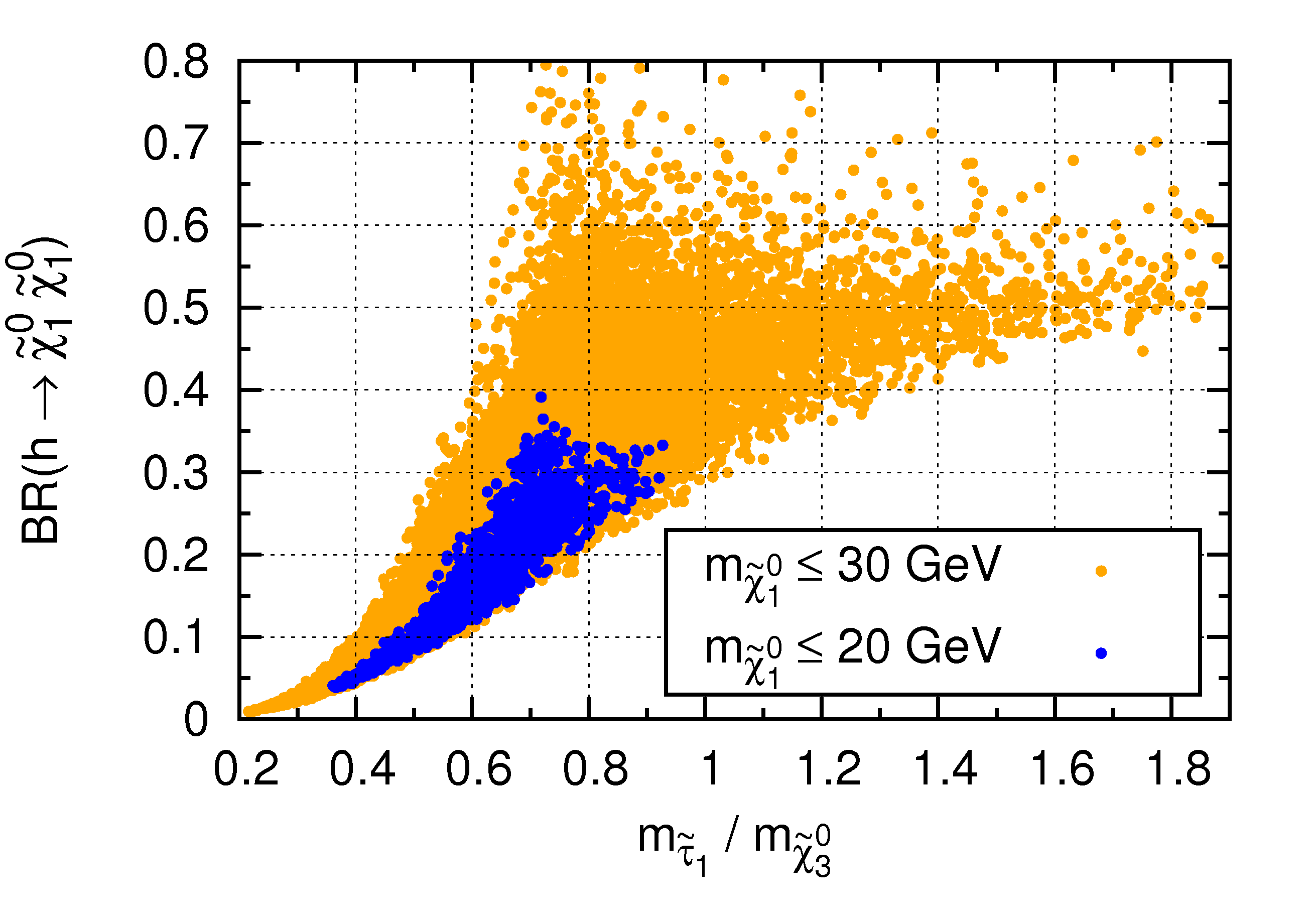}
 \includegraphics[width=0.45\textwidth]{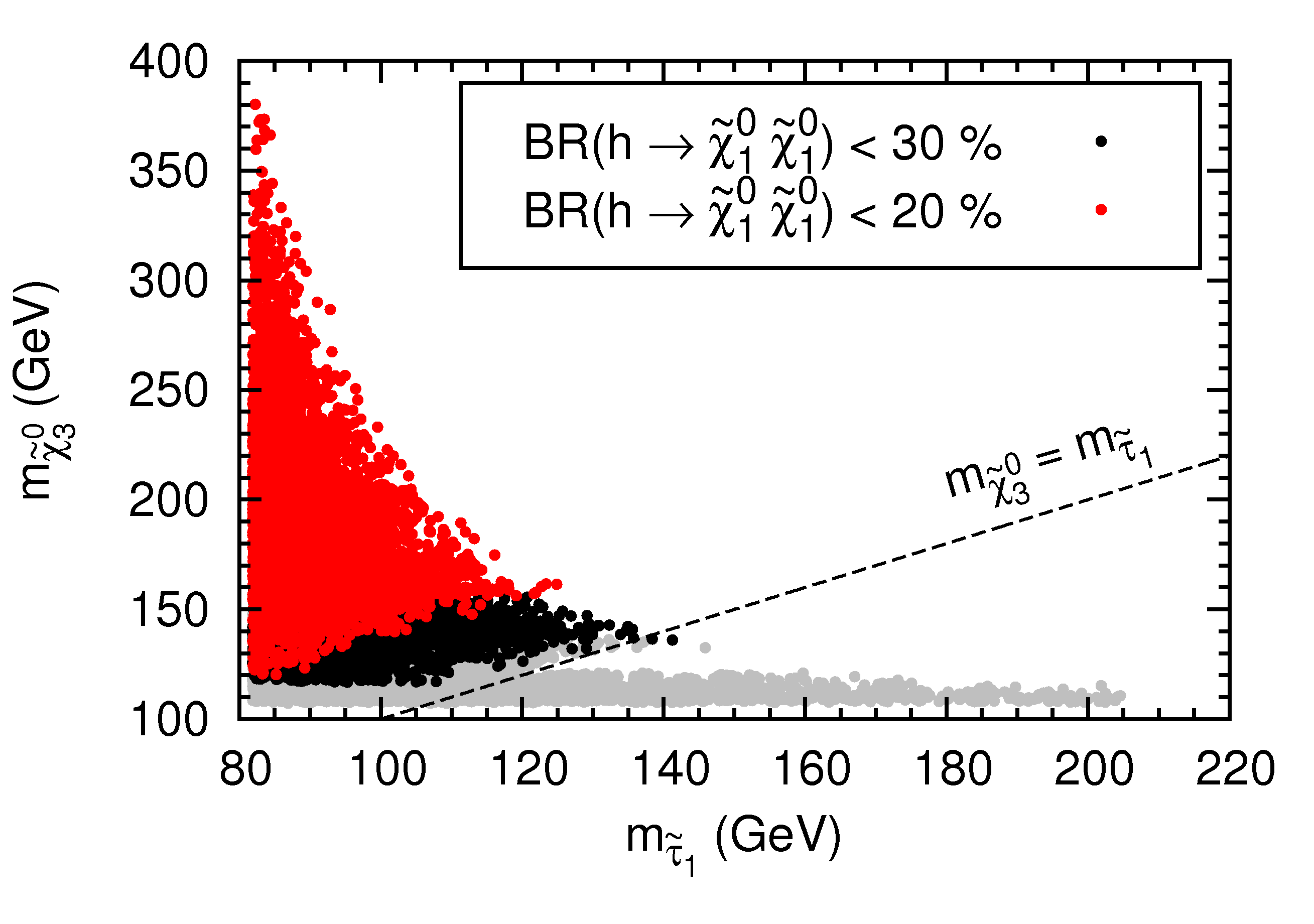}
 \caption{Left: BR($h\to \nt_1\nt_1$) versus the mass ratio
   $m_{\stau_1}/m_{\nt_3}$, all the constraints of section
   \ref{sec:relicdensity} are applied. Right: different values for
   BR($h\to \nt_1\nt_1$) displayed on the same
   $m_{\stau_1} - m_{\chi_3}$ plane of Fig.~\ref{fig:stau-chi}.
\label{fig:hchichi} }
\end{figure}
As discussed in section \ref{sec:parspace}, relic density constraints
require sizeable Higgsino components in $\nt_1$ that can induce large
branching ratios for the decay $h\to
\nt_1\nt_1$~\cite{Griest:1987qv,Yaguna:2007vm,Dreiner:2012ex}.  This
is in particular the case for heavier $\stau_1$ and thus for the
hierarchy $m_{\stau_1} > m_{\nt_{2,3}} > m_{\nt_1}$. Here, fulfilling
the relic density bound requires a larger Higgsino component (small
$\mu$) to contribute to $\nt_1$ annihilation.  The light Higgs decay
width into the lightest neutralinos is given by~\cite{Griest:1987qv}:
\begin{equation}
\Gamma (h \rightarrow \nt_1 \nt_1)= \frac{G_F M_W^2 m_h}{2 \sqrt{2} \pi}~
\left(  1- \frac{4 m_{\nt_1 }^2}{m_h^2}   \right)^{3/2} 
\big\vert  C_{h \nt_1 \nt_1 }   \big\vert^2 \;,
\end{equation}
where in the decoupling regime $m_A \gg m_h$
\begin{equation}
 C_{h \nt_1 \nt_1 } = \big( N_{12} -\tan \theta_W \; N_{11}   \big)
\big( \sin\beta\; N_{14} -\cos\beta \; N_{13}   \big)\;.
\end{equation}
Thus, we see that, in contrast to $\Gamma(Z\to \nt_1 \nt_1)$, the
contribution from the ${\widetilde H}_u$ component is dominant as it
is $\tan\beta$ enhanced with respect to the ${\widetilde H}_d$ one.

In Fig.~\ref{fig:hchichi} (left) we plot BR($h\to \nt_1\nt_1$) versus
$m_{\stau_1}/m_{\nt_3}$ using the parameter scan illustrated
in section \ref{sec:parspace}. BR($h\to \nt_1\nt_1$) has been computed using
{\tt SUSY-HIT}~\cite{Djouadi:2006bz}.  We find
that amost all the allowed points with $m_{\stau_{1}} /m_{\nt_{3}} > 1$
(i.e. the points below the dashed-line in the right plot of 
Fig.~\ref{fig:hchichi})
correspond to a branching ratio larger than 30\%.
This is to be compared to the constraints from fits of 
${\rm BR}_h^{\rm inv}\equiv{\rm BR}(h \to {\rm invisible})$ to the observed
Higgs decay rates~\cite{Giardino:2013bma,Falkowski:2013dza}:
\begin{align}
 {\rm BR}_h^{\rm inv} \lesssim 20 \%\quad(95\%~{\rm CL}). 
\label{eq:inv}
\end{align}
{}From this we see that the case $m_{\stau_1} > m_{\nt_{2,3}}$ is
strongly disfavored, while $m_{\nt_{2,3}} > m_{\stau_1}$ is still
viable but partly constrained, as shown in Fig.~\ref{fig:hchichi}
(right).\footnote{See, however, the more conservative bound obtained
  in Ref.~\cite{Djouadi:2013qya}, considering large theoretical
  uncertainties: ${\rm BR}_h^{\rm inv} \lesssim 52 \%~(68\%~{\rm CL})$.}

\subsection{Monojet searches}	
\label{sec:monojet}
%%%%%%%%%%%%%%%%%%%%%%%%%%%%%%%%%%%%%%%%%%%%%%%%%%%%%%%
\begin{figure}
\centering
 \includegraphics[width=0.45\textwidth]{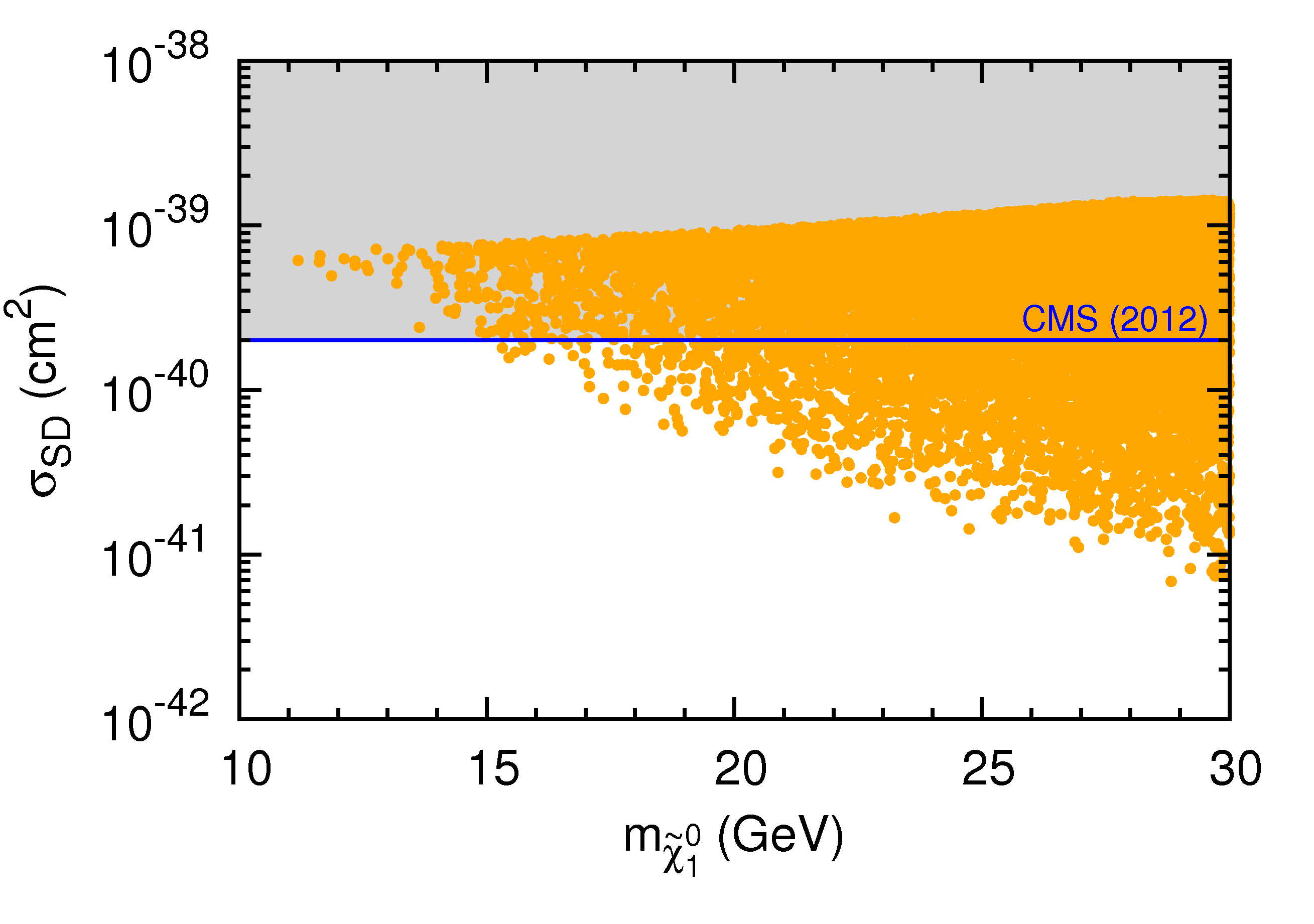}
 \includegraphics[width=0.45\textwidth]{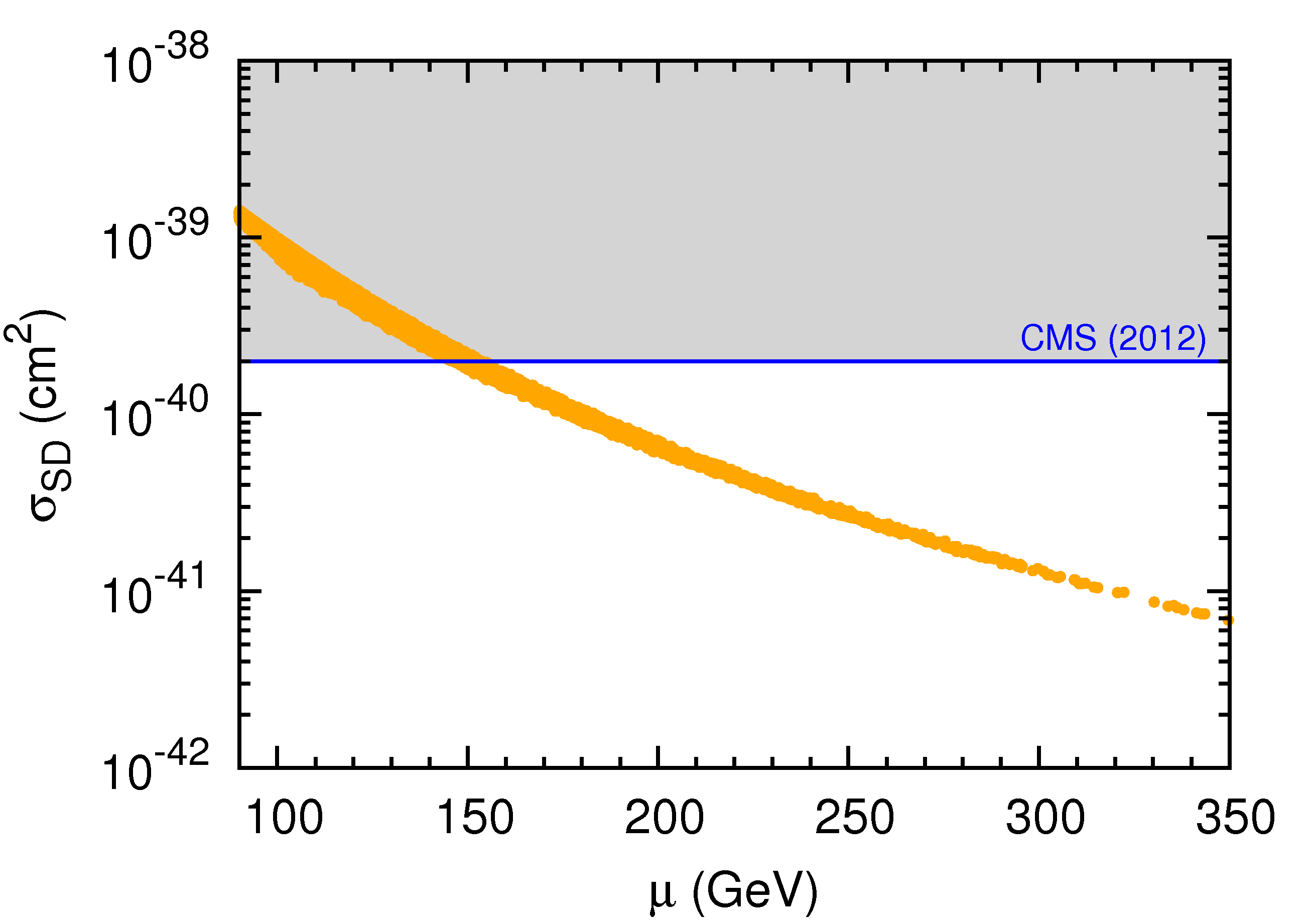}
 \caption{Model prediction for the SD scattering proton-neutralino
   cross-section as a function of the neutralino mass (left) and the
   Higgsino mass parameter $\mu$ (right), compared to the current
   bound from monojet searches at the LHC.\label{fig:monojet} }
\end{figure}
%%%%%%%%%%%%%%%%%%%%%%%%%%%%%%%%%%%%%%%%%%%%%%%%%%%%%%%
%
As anticipated in section \ref{sec:DMsearches}, direct DM searches at
the LHC, based on monojet + $\missingET$ events, can be displayed as
limits on the neutralino scattering cross-section with nuclei. As we
discussed above, the prediction for the spin-dependent cross-section,
$\sigma_{\rm SD}$, is not too sensitive to the SUSY spectrum under
consideration.  Indeed, the effective proton-neutralino interaction is
mediated by a $Z$ boson and hence is solely determined by the size of
the Higgsino components in $\nt_1$.  We expect that the larger the
components are -- corresponding to small values of $\mu$ --, the
stronger the bound from LHC searches becomes.  We computed $\sigma_{\rm
  SD}$ by means of {\tt micrOMEGAs}, checking that variations of the
hadronic matrix elements affect the results only at the level of
$10\div 12$~\%.%\footnote{Such theoretical uncertainties are taken into
%   account at the end of the next section where we summarize our
%   results.}
In the interpretation of the 
ATLAS~\cite{ATLAS:2012ky} and CMS~\cite{Chatrchyan:2012me} analyses 
--
the latter giving a slightly stronger limit 
--
in terms of $\sigma_{\rm SD}$, 
we take into account the fact that neutralinos 
are Majorana particles and thus the published
limits are in our case weaker by a phase-space factor of
two~\cite{Goodman:2010yf,Goodman:2010ku}.  

The result is shown in Fig.~\ref{fig:monojet}, from which we see that
the parameter space is considerably constrained by the monojet data.
In particular, the lower bound of the neutralino mass (left panel) is
raised to
\begin{equation}
m_{\nt_1} \gtrsim 15~\text{GeV}
\end{equation}
and points with low values of the Higgsino mass -- namely $\mu
\lesssim 150$ GeV -- are excluded (right panel).
As in the case of the invisible Higgs decay discussed above, this
observable is therefore complementary to the multi-tau searches in
testing the light neutralino scenario: in fact, the parameter region
with $m_{\stau_1} > m_{\nt_{2,3}}$, which is kinematically
unfavorable for the multi-tau searches, requires $\mu \approx 100\div
130$ GeV (cf. Fig.~\ref{fig:stau-chi}).
{}From the comparison between Fig.~\ref{fig:stau-chi} and the left panel
of Fig.~\ref{fig:monojet}, it is reasonable to conclude that the null
result of the monojet searches at the LHC would not be compatible with the
parameter region of the light neutralino dark matter scenario with
$m_{\stau_1} > m_{\nt_{2,3}}$.
However, for a conclusive statement of the potential of the LHC monojet searches on direct neutralino production and its relation with direct dark matter detection limits, a detailed study beyond the effective field theory approximation\footnote{For a recent study about the limits of such an approximation, we refer to \cite{Buchmueller:2013dya}.} carried out here is necessary.

\section{LHC multi-tau limits}
\label{sec:LHC_numerics}

\subsection{ATLAS multi-tau analysis and Monte Carlo framework}

Recently the ATLAS collaboration presented a (preliminary) analysis of
a search for new physics in a final state with multi-taus and large
missing transverse energy \cite{ATLAS:2013yla} employing a data sample
of $20~\fbai$ at $\sqrt{s}=8$ TeV.
In Ref.~\cite{ATLAS:2013yla} at least
two reconstructed hadronic taus are required together with a missing
transverse energy of $\missingET > 40$ GeV; the final event selection
and background suppression are based on a cut on $m_{T2}$
\cite{Lester:1999tx,Barr:2003rg} and different jet vetoes. Resulting
limits are presented in different simplified models and a parameter
region of the ``phenomenological MSSM'' (pMSSM) with light charginos, 
neutralinos and staus. The presented analysis is inclusive in the number 
of reconstructed hadronic taus and relevant for the light neutralino scenario
investigated in this study. In the following we reproduce the pMSSM
limits presented in Ref.~\cite{ATLAS:2013yla} and reinterpret them in the
light neutralino parameter space discussed above.

For the signal event simulation we use \herwig\ \cite{Bahr:2008pv} and
include production of neutralinos, charginos and sleptons. For any
scenario we consider, squarks and gluinos are assumed to be heavy and their
productions do not contribute.
Everywhere full spin correlations in the decays,
initial-state-radiation (ISR), final-state-radiation (FSR),
hadronization effects, and underlying-event-simulation are
included. Obtained event samples are normalized to inclusive NLO 
cross-sections calculated with \prospino\ \cite{Beenakker:1999xh}.
Via the \hepmc\ format \cite{Dobbs:2001ck} signal events are passed to
\delphes\ \cite{Ovyn:2009tx} for (fast) detector simulation and event
reconstruction.  We use the default ATLAS detector card within
\delphes\ and tune all efficiencies to the values given in
Ref.~\cite{ATLAS:2013yla}. Signal events are reconstructed and
selected, as in Ref.~\cite{ATLAS:2013yla}, according to the following
criteria.
Light central jets are required to have 
transverse momentum $\pT_j > 25~\GeV$ and
pseudorapidity $\vert\eta_j\vert<2.5$. 
Forward jets must satisfy $\pT_{j_f} >
30~\GeV$ and $2.5< \vert\eta_{j_f}\vert<4.5$. For tagged b-jets we
require $\pT_{j_b} > 20~\GeV$ and $\vert\eta_{j_b}\vert<2.5$. For
electrons and muons we require $\pT_l > 10~\GeV$, and
$\vert\eta_e\vert<2.47$ and $\vert\eta_{\mu}\vert<2.4$, respectively.
Hadronic tau candidates are required to have $\pT_{\tau_h} > 20~\GeV$
and $\vert\eta_j\vert<2.5$. Now we select events with at least two
opposite sign (OS) taus. Additional light leptons are vetoed and we
require $\missingET > 40~\GeV$.  The variable $m_{T2}$ is calculated
using \cite{oxbridge} and the assumed mass of the neutralino is set to
$m_{\text{inv}}=0$. In events where more than two taus are
reconstructed, $m_{T2}$ is computed taking all possible OS tau pairs
into account and then choosing the largest value. For the final event
selection two signal regions are defined as in Ref.~\cite{ATLAS:2013yla}:
\begin{itemize}
 \item SR1: veto of central light jets and forward jets, and a cut on $m_{T2}>90\gev$,
 \item SR2: veto of b-jets, and a cut on $m_{T2}>100\gev$.
\end{itemize}

In these signal regions the analysis presented in Ref.~\cite{ATLAS:2013yla}
sets at $95\%$ CL the following limits on the number of signal events:
\begin{align}
 S^{95}_{\rm SR1} < 5.6  \hspace{1cm} \text{and}\hspace{1cm}  S^{95}_{\rm SR2}< 10.4 \, .
\end{align}
These limits are then interpreted in an $M_2-\mu$ plane of the pMSSM
with $\tan\beta=50$, $M_1=50~\GeV$ and $m_{\tilde\tau_R}=84.7~\GeV$
(together with $A_{\tau}=5~\TeV$ this yields a lighter stau mass of
$m_{\tilde\tau_1}\approx 95~\GeV$). All other parameters are
decoupled.
\begin{figure}[t]
\centering
\includegraphics[width=.6\textwidth]{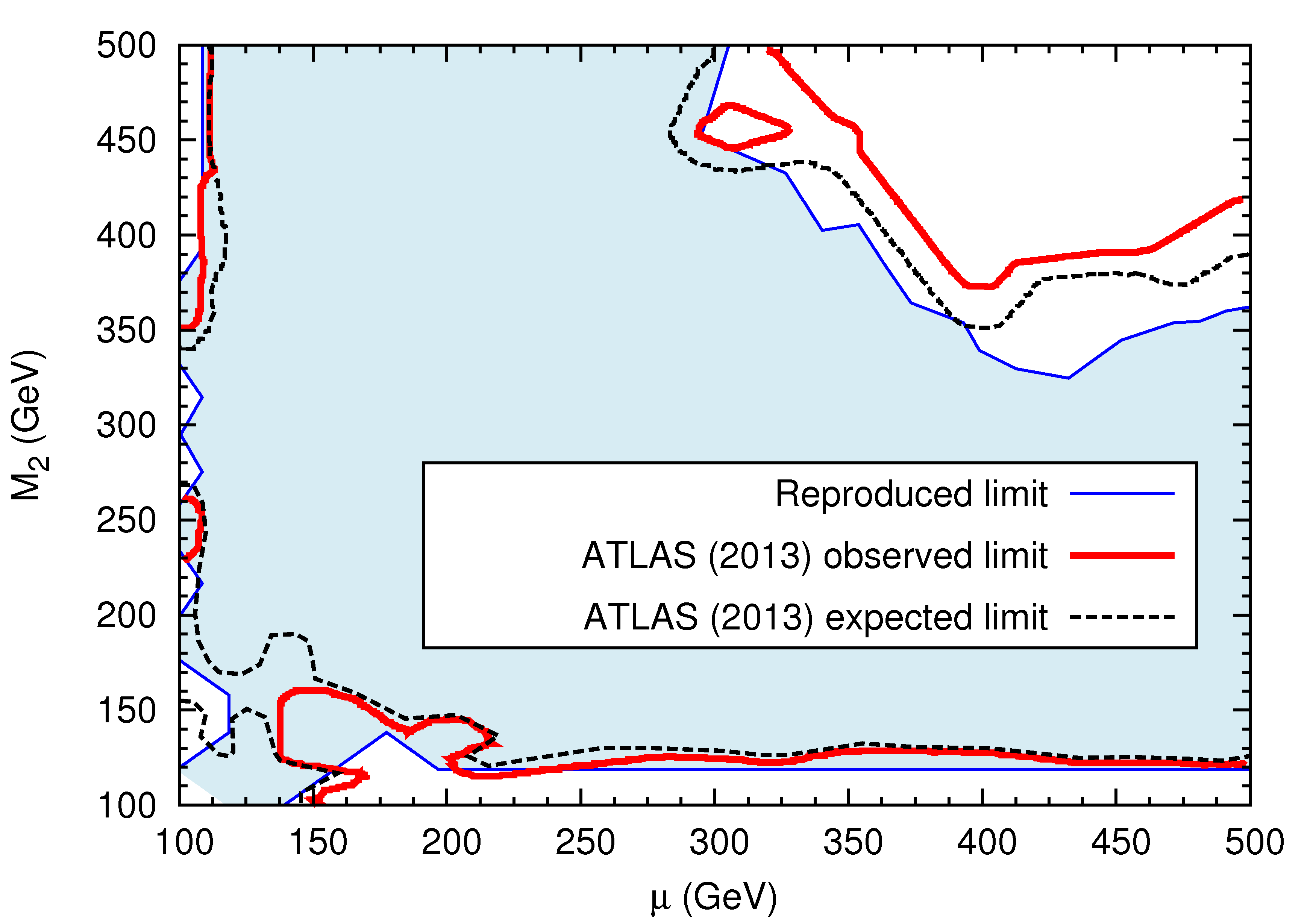} 
\caption{Excluded regions in the $M_{2}-\mu$ parameter plane of the
  pMSSM. The bounds obtained by \cite{ATLAS:2013yla} and the
  limits reproduced by our simulation are shown.\label{fig:atlas_exclusion}}
\end{figure}

With our Monte Carlo framework we perform a scan in the very same
pMSSM plane and compare the resulting number of signal events with the
above stated limits at $95\%$ CL. The result is shown in
Fig.~\ref{fig:atlas_exclusion}. Besides the exclusions obtained with
our framework we show the expected and observed exclusions of
Ref.~\cite{ATLAS:2013yla}. The agreement between the reproduced
exclusions and the ones given in Ref.~\cite{ATLAS:2013yla} seems to be
sufficient in order to allow a reinterpretation of the underlying
limits in the light neutralino parameter space.

\subsection{Interpretation for the light neutralino scenario}
\begin{figure}[t]
\centering
\includegraphics[width=.6\textwidth]{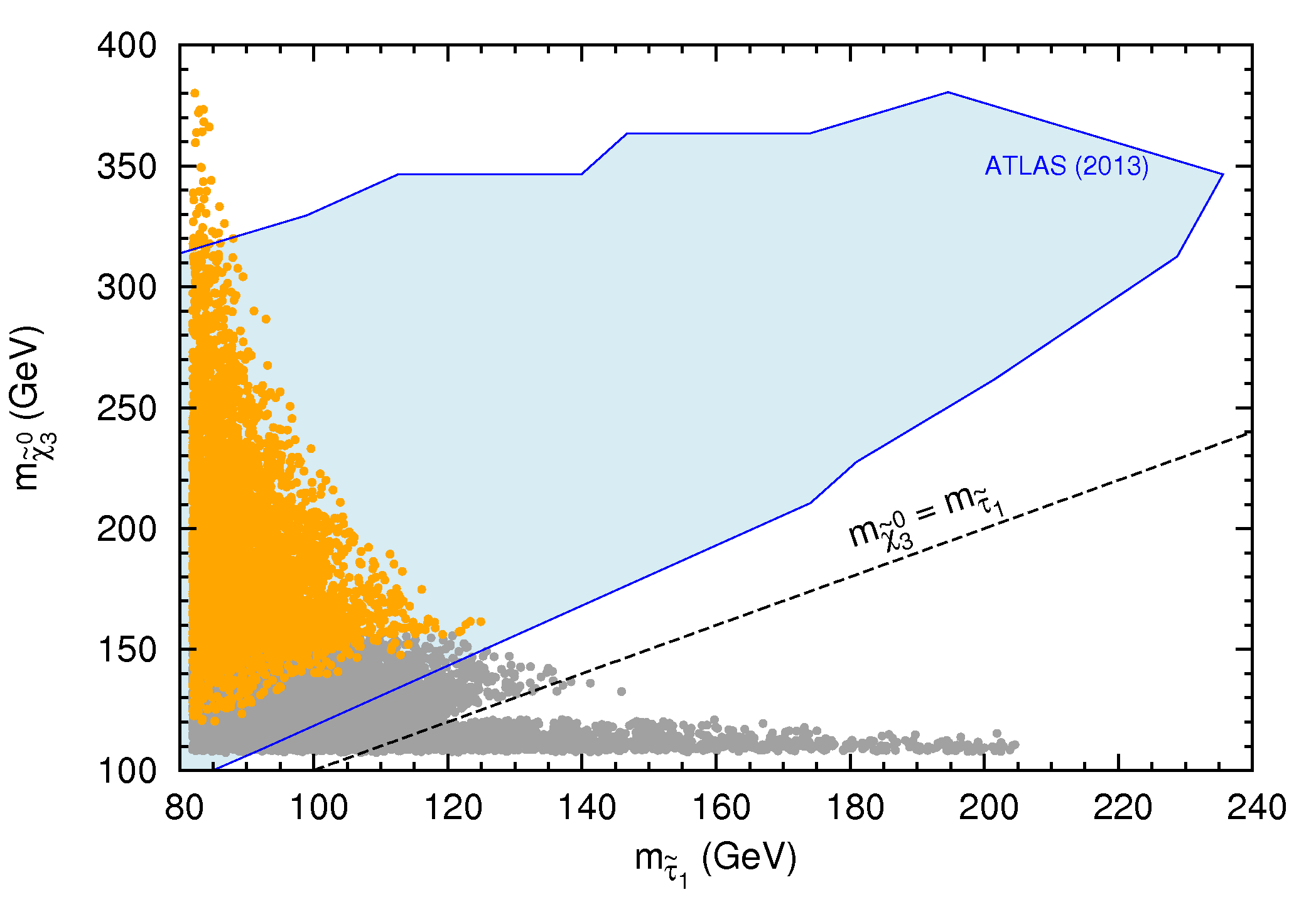} 
\caption{Region excluded by ATLAS in the $m_{\tilde\tau_1}-m_{\nt_3}$
  parameter plane. The light neutralino parameter space is also shown:
  the gray points fulfill the constraints of section
  \ref{sec:relicdensity}, the orange points additionally 
 give ${\rm BR}(h\to \nt_1\nt_1)< 20\%$, see section \ref{sec:invH}.\label{fig:our_exclusion}}
\end{figure}
\begin{figure}[t]
\centering
\includegraphics[width=.6\textwidth]{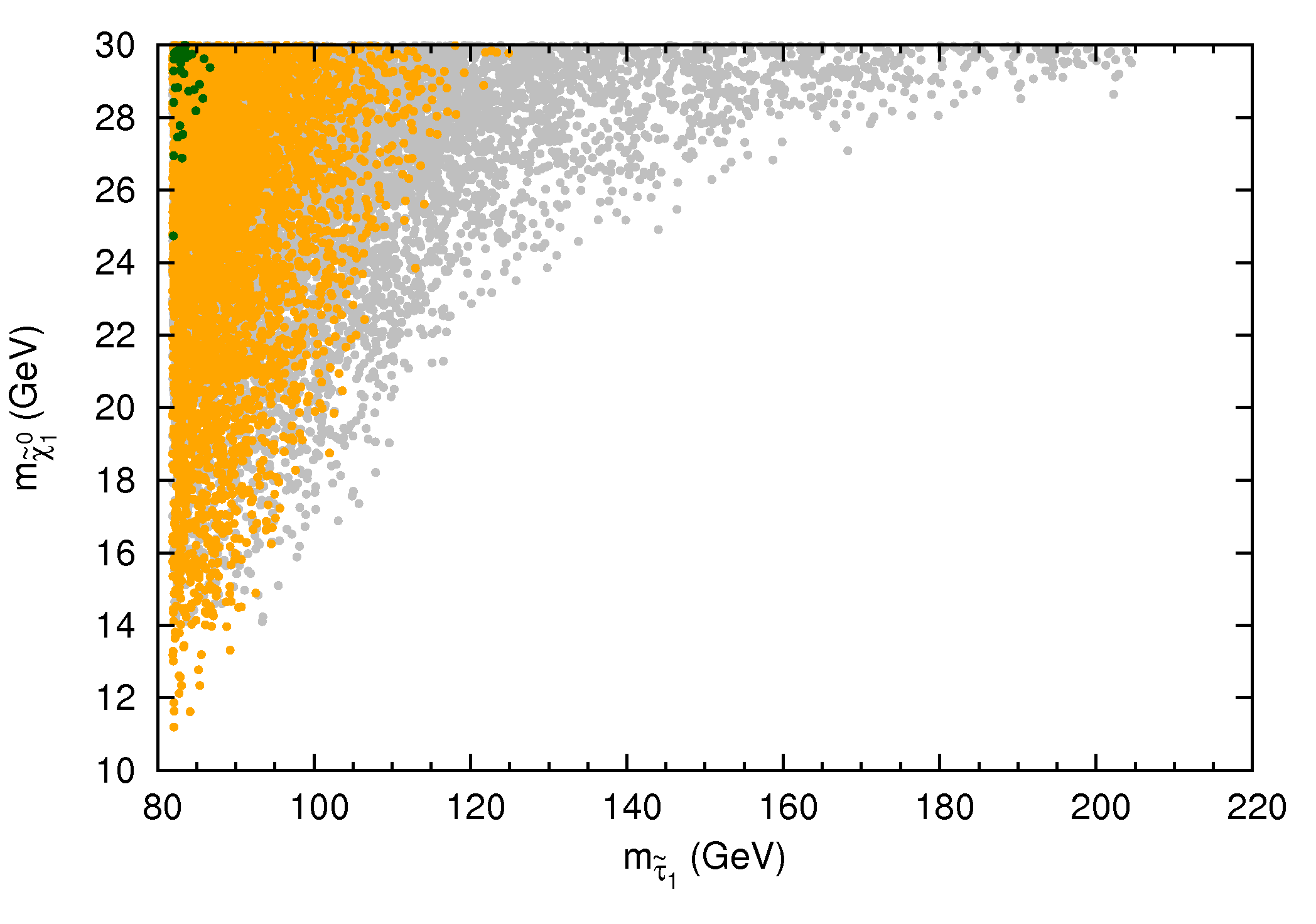} 
\caption{Summary of the LHC searches in $m_{\tilde\tau_1}-m_{\nt_1}$
  plane: the gray points fulfill the constraints of section
  \ref{sec:relicdensity}, the orange points correspond to ${\rm BR}(h\to \nt_1\nt_1)< 20\%$, 
  the dark-green points evade the ATLAS multi-tau limit of
  Fig.~\ref{fig:our_exclusion}.\label{fig:our_exclusion2}}
\end{figure}
As discussed in detail in section \ref{sec:LHC_pheno}, we expect
striking multi-tau signals for the considered light neutralino
parameter space, and consequently possibly strong constraints on this
parameter space from the ATLAS analysis in
Ref.~\cite{ATLAS:2013yla}. In Fig.~\ref{fig:our_exclusion} we
reinterpret the limits of \cite{ATLAS:2013yla} in the
$m_{\tilde\tau_1}-m_{\nt_3}$ plane discussed above.  In this
reinterpretation, we set $\tan\beta=55$, $M_1=30~\GeV$, and as before
all other mass parameters with the exception of the relevant stau and
Higgsino masses are assumed to be heavy.  We checked that limits for
an even lighter neutralino are at least as strong as the obtained
ones. Moreover, the limits obtained for our simplified model do hardly
depend on the choice of $\tan\beta$.  From the figure, we see that
parameter regions with $m_{\nt_3}>m_{\tilde\tau_1}$, where the heavier
neutralinos can decay into on-shell lighter staus are excluded up to
$m_{\ch_1} \approx m_{\nt_{2,3}}\gtrsim 320~\GeV$. For
$m_{\nt_3}\lesssim m_{\tilde\tau_1}$ decays of the heavier neutralinos
into taus are only possible via off-shell decays, where various decay
modes compete and corresponding exclusions limits are much
weaker. However, as discussed in section \ref{sec:invH} and
\ref{sec:monojet}, such parameter regions are strongly disfavored by
recent limits on the invisible width of the light Higgs, $h$, and potentially by monojet searches at the LHC.
In Fig.~\ref{fig:our_exclusion} the orange points 
give ${\rm BR}(h\to \nt_1\nt_1)< 20\%$ (cf.~the right panel of Fig.~\ref{fig:stau-chi}),
while the gray points
fulfill the relic density constraints (and the other constrains
introduced in section \ref{sec:relicdensity}). 
Additionally, in the region where $m_{\nt_{2,3}} - m_{\nt_1} >
m_{Z}$ limits from dedicated searches in final states with SM gauge
bosons and missing transverse energy could become relevant
\cite{Aad:2012hba,Chatrchyan:2012pka}. Still, in the light neutralino
scenario this is a very small parameter region.  Multi-tau searches
with higher luminosity and center-of-mass energy $\sqrt{s}$, in
combination with monojet searches and Higgs decay measurements, are
the most promising way to test light neutralino DM up to $m_{\nt_1}
\approx 30$ GeV. Indeed, only a tiny corner of the parameter space is
left unprobed by the different experimental information discussed
above: this is better depicted in Fig.~\ref{fig:our_exclusion2}, where
we show the impact of the ATLAS limit on the $m_{\stau_1}-m_{\nt_1}$
plane.  As in Fig.~\ref{fig:our_exclusion}, the gray points fulfill
the constraints discussed in section \ref{sec:relicdensity} and the
orange points additionally 
correspond to ${\rm BR}(h\to \nt_1\nt_1)< 20\%$.
The dark-green
points are the only ones left after the ATLAS multi-tau search.  This
allows us to set the present lower bound on the neutralino mass --
assuming Eqs.~(\ref{eq:planck}) and (\ref{eq:inv}) -- at about
\begin{equation}
m_{\widetilde{\chi}^{0}_{1}}  > 24 \div 25~\text{GeV}.
\label{eq:final-mass-limit}
\end{equation}
In addition, we see that the few points left require a very light stau
with $m_{\stau_1}\lesssim 90$ GeV, a value very close to the LEP
exclusion limit reported in Ref.~\cite{PDG}.

\section{Summary and conclusions}
\label{sec:conclusions}

In this paper, we have shown that, assuming a light neutralino as
DM candidate and the particle content of the MSSM, the bounds from the
DM relic abundance can be fulfilled only in a limited and well-defined region 
of the parameter space. For a mass of the lightest neutralino smaller than about 30 GeV
a handful of parameters suffices to define this parameter space. 
In turn, this allows to employ current searches for SUSY at the LHC 
to set definite limits on light neutralino DM and give a lower bound on its mass.

The allowed region of the parameter space is characterized by
a relatively light right-handed stau, a light Higgsino-dominated chargino and
light Higgsino-dominated neutralinos, cf. Eq.~(\ref{eq:up-bounds}). 
Therefore, the recent search for multi-tau $+$ $\missingET$ events performed 
by the ATLAS collaboration \cite{ATLAS:2013yla} can be interpreted as a test of the light
neutralino DM scenario. 
By means of a Monte Carlo simulation including fast detector simulation, 
we have shown how this ATLAS analysis strongly constrains the relevant 
parameter space. We have also highlighted the complementarity of such 
tests with other new physics observables at the LHC like the search for the 
decay $h\to{~\rm invisible}$, or searches in the monojet $+$ $\missingET$ 
channel. 
In combination with this experimental information, the multi-tau 
ATLAS search excludes most of the parameter space with 
$m_{\nt_1}< 30$ GeV, setting a lower bound on the
neutralino DM mass: $m_{\widetilde{\chi}^{0}_{1}} > 24 \div
25~\text{GeV}$.  
The plots in Figs.~\ref{fig:our_exclusion} and
\ref{fig:our_exclusion2} present our final results and show how tiny the 
remaining parameter space is.
Clearly, a small increase of the sensitivity in this
channel at the future $\sqrt{s}=13\div14$ TeV run of the LHC can 
completely test the light neutralino DM scenario up to $m_{\nt_1} \approx 30$ GeV.

Larger values of the lightest neutralino mass cannot be probed in such a
unique way: in fact, as we argued in section \ref{sec:lightDM},
$m_{\nt_1} > 30$ GeV would open the possibility of satisfying the
relic density constraints with compressed spectra that can, at the
same time, (i) evade the LEP searches for light sfermions, (ii) be
insensitive to constraints from $Z$-pole observables, (iii) be
very hard to be tested at the LHC. Furthermore, for even larger
$m_{\nt_1}$, a very efficient neutralino annihilation would be
possible through resonant $Z$ and/or $h$ exchange that would just
require non-vanishing Higgsino components in $\nt_1$. A detailed
discussion of such possibilities is beyond the scope of this paper and
will be given elsewhere.

A crucial assumption leading to the above results is that 
the lightest neutralino constitutes (mainly) the observed DM in the
universe.
Dropping such a hypothesis, as for instance in models with (small)
R-parity violation, would clearly change completely the discussion on
how light the neutralino can be. 
Corresponding phenomenology at colliders could be very different, e.g. in the case of a
promptly decaying neutralino.  For early discussions on (very) light
neutralinos without cosmological bounds, we refer to Refs.
\cite{Dreiner:2009ic,Dreiner:2009er} and references therein.

Concerning direct underground DM searches, we have shown in section
\ref{sec:DMsearches} that the scattering cross-sections with nuclei
predicted in the present scenario are in the range $10^{-43}~{\rm cm^2}
\lesssim \sigma_{\rm SI}\lesssim 10^{-46}~{\rm cm^2}$.  This implies
that (i) light MSSM neutralinos cannot account for the signals
reported by several direct detection experiments, (ii) the scenario we
discussed has cross-sections close to the current XENON100 sensitivity and will be
fully tested independently by XENON1T (cf.~the left panel of
Fig.~\ref{fig:DMsearches}).  
Similar conclusions can be drawn for
indirect searches: the light neutralino scenario lies close to the border of
the current Fermi-LAT sensitivity. Thus, it might be complementary 
tested in the near future also by gamma-ray observations 
(cf. Fig.~\ref{fig:DMsearches}, right).

To conclude, let us remark that the presented study provides an example
of the amazing capability of the LHC experiments to test new
physics through pure electroweak interactions, as well as of the
complementarity of different collider searches and other experimental information
(CMB observations, direct/indirect DM searches, etc.) in shedding
light on nature and properties of dark matter and physics beyond the Standard Model
in general.

\section*{Acknowledgments}
\noindent We thank Wolfgang Hollik for providing us with
the numerical implementation of Ref.~\cite{Heinemeyer:2007bw}. 
We would also like to thank Federica Legger, Borge Gjelsten and Xuai Zhuang for explanation of 
various details of Ref.~\cite{ATLAS:2013yla}. Furthermore, we acknowledge 
helpful comments by Herbi Dreiner on the development of the light 
neutralino scenario and discussions with Michael Gustafsson about indirect DM searches. 
TO thanks Carlos Yaguna for his insightful comments on the invisible Higgs decay.
LC, TO, and YT are grateful to the Max-Planck-Institut f\"{u}r Physik for
the support and hospitality during their visit.
%%%%%%%%%%%%%%%%%%%%%%%%%%%

\end{document}